\documentclass[12pt]{article}
\usepackage{amsmath,amssymb,bm}
\usepackage{graphicx}
\usepackage{enumerate}
\usepackage{natbib}
\usepackage{url} 

\usepackage{array}

\usepackage{enumitem}
\usepackage{amsthm}

\newcommand{\blind}{0}

\newtheorem{theorem}{Theorem}

\renewcommand\u{\underline}
\newcommand\bs{\boldsymbol}
\newcommand\bmu{\bs\mu}
\newcommand\bX{\bs X}
\newcommand\bSigma{\bs\Sigma}
\renewcommand\mod{{\rm mod}}
\newcommand\ba{\begin{array}}
\newcommand\ea{\end{array}}
\newcommand\ms{\medskip}
\newcommand\st{\buildrel {st}\over\sim}
\newcommand\equivd{\buildrel {d}\over\equiv}
\newcommand\as{\buildrel a\over\sim}
\newcommand\ds{\displaystyle}
\newcommand\be{\begin{equation}}
\newcommand\ee{\end{equation}}
\newcommand\nref[1]{(\ref{#1})}
\newcommand\ov{\overline}
\newcommand\by{{\scriptstyle\times}}
\newcommand\ii{{\rm i}}

\begin{document}

\def\spacingset#1{\renewcommand{\baselinestretch}%
{#1}\small\normalsize} 
\spacingset{1}


  \title{\bf A robust Likelihood Ratio Test for\\[4pt] high-dimensional MANOVA -- with excellent performance}
  \author{\LARGE{Carlos A. Coelho}
\vspace{.2cm}\\
Department of Mathematics and NOVA-Math\\NOVA School of Science and Techonolgy, NOVA University of Lisbon}
		\date{}
  \maketitle

\if0\blind
{
  \bigskip
  \bigskip
  \bigskip
  \medskip
} \fi

\bigskip
\begin{abstract}
The present paper answers the following questions related with high-dimensional {\sc manova}: (i) is it possible to develop a likelihood ratio test for high-dimensional {\sc manova}? (ii) would such test perform well? (iii) would it be able to outperform existing tests? (iv) would it be applicable to extremely small samples? (v) would it be applicable to non-normal random variables, as uniform, extremely skewed distributions, or even heavy tailed distributions with success? (vi) would it have a nice, rather simple to compute and well performing, asymptotic distribution? And what about if the answer to all the above questions would be a clear ``Yes''? Surprisingly enough, it is exactly the case. Extensive simulations, with both normal and non-normal distributions, some of which are heavy tailed and/or highly skewed, and even discrete distributions, are carried out in order to evaluate the performance of the proposed test and to compare its performance with other tests. Two real data applications are presented.
\end{abstract}

\noindent%

\section{Introduction}
\label{sec:intro}

\subsection{The problem and the objectives}

With the advent of faster and less expensive computers and tools and physical supports to collect and store large amounts of data, arises the need for statistical methods and techniques able to handle such large amounts of data. 
In this paper we are concerned with a particular type of such data, that is, situations in which the number of variables measured is much larger than the number of observational units, and namely situations where the sample sizes may be really extremely small. We will call these situations as high-dimensional, as it is actually usually done.
With the development of measuring devices capable of easily measuring a large number of variables on a single observational unit or station, the advent of such types of data is more and more common.
These are situations that are nowadays common in many research areas, as chemometrics \citep{Perouetal2000}, medicine, metabolomics \citep{Brinietal2021,Condeetal2022}, proteomics \citep{Moumetal2015,Gomesetal2019} and other omics, as in genomics and genetic studies \citep{Perouetal1999,WangZhu2008,Ilketal2011,Chenetal2015,Lietal2019,Konetal2021}. But such situations may also happen in many other areas of research as for example in social sciences \citep{Solomonetal12} or econometric studies \citep{Kwon2007,Fanetal11,Belloni11,Bellonietal14}, being the case that \citet{Kao19} even state in the presentation of their book that ``High-dimensional data in econometrics is the rule rather than the exception'', as well as in areas as agronomy \citep{Chavesetal2009,Figueiredoetal2017,Neilsonetal2017}, forestry and seismology \citep{RostThomas2002},
where a large number of variables or characteristics is measured on a small number of available observational units or individuals.

Commonly these observational units are divided in different groups or populations, which may be determined by gender, race, educational background, marital status, diagnostic, geographical origin or geographical region, different agronomical varieties or forest subspecies, etc.

The present paper is concerned with the development of a testing procedure adequate to test for differences among the mean vectors, of the $p$ variables measured, for the several populations involved in the study. Furthermore, it is an objective that such procedure may be applied in situations where the underlying random variables (r.v.'s) do not necessarily follow a Normal distribution, but rather may have almost any distribution, and also in situations where the samples may be really small, compared with the number of variables involved, preferably even allowing the samples in some of the populations to have just size 1, while assuring that the test procedure  keeps a `good performance'.

In the seminal papers by \citet{ChungFraser} and \citet{Dempster,Dempster2} two quite different approaches were used in developing tests for the equality of two mean vectors in the high-dimensional case. While the first authors used a randomization and permutation based technique, Dempster introduced an Euclidean metric on the space of variables that allowed him to define a
test statistic that is the ratio of two mean square distances, this way avoiding the problem that appears when we may think about using Hotelling's $T^2$ statistic in the high-dimensional case. Dempster actually uses a data transformation to orthogonal vectors, a technique somewhat related to random projection techniques, later used by other authors. However, while randomization and permutation based tests also have problems when dealing with very small samples, since their p-values rely on the number of possible permutations, and as such, for example for two samples with sizes $n_1=n_2=4$, the lowest attainable p-value would be $1/(8!/(4!4!))=0.014$, on the other hand Depmster's approach debouches into a rather elaborate estimation procedure of a distribution parameter.

Indeed, a natural test statistic for the difference of two population mean vectors may seem to be one based on the difference between the two sample mean vectors, weighted by the inverse of the pooled sample covariance matrix. This is actually an estimator of the Mahalanobis distance \citep{Mahalanobis} between the two population mean vectors, and it is also the Hotelling $T^2$ \citep{Hotelling} statistic to test the equality of two population mean vectors in the non high-dimensional setting. But, in the high-dimensional case the sample pooled covariance matrix is not invertible, since such inverse only exists when $n-1>p$. It was exactly to avoid this problem that several lines of research emerged.

One of them was
centered around developing test statistics that, although keeping similarities with the $T^2$ Hotelling statistic, being based on a measure of distance between the two sample mean vectors, do not use the inverse of the pooled covariance matrix. \citet{BaiSaranadasa96} replaced the pooled sample covariance matrix by an identity matrix when computing the 
distance between the two sample mean vectors and then subtract from this distance a quantity proportional to the trace of this pooled covariance matrix, to make their test statistic to have the an expected value equal to the square norm of the difference of the two population mean vectors, while \citet{SrivDu08} proposed to replace the pooled sample covariance
 matrix by its diagonal, and \citet{Sriv07} replaced the inverse of the pooled sample covariance matrix by its Moore-Penrose inverse. 
\citet{ChenQin10} took a test statistic based on the squared norm of the difference between the two sample mean vectors, from which they subtract the cross-product terms. They argue that by doing this not only the expected value of the test statistic will be equal to the squared norm of the difference between the two population mean vectors but also it avoids the need for a restriction relating the larger eigenvalue of the population covariance matrix and its dimension. But this non-inclusion in the computation of the test statistic of the cross-product terms actually, besides making the computations harder, also makes the test to behave quite badly for most non-normal distributions, as we will see in section 4.
\citet{Zhangetal2020} propose a test based on a statistic similar to the one proposed by \citet{BaiSaranadasa96}, without including the subtraction of the trace of the pooled sample covariance matrix, but which then
incorporates in a interesting way functions of this covariance matrix in the computation of the distribution of the test statistic, which is approximated by a Gamma distribution. Moreover, these authors also provide what they claim to be, and indeed it is, an improvement in the estimation of the distribution of the test statistic for non-normal populations. Some further details on this test may be seen in Subsection 1.4.. \citet{Fujietal04} and \citet{Schott07} proposed a test for $q>2$ populations which use statistics based on the traces of the 
within and between sample covariance matrices, the same that are actually used in the LRT for the test of equality of mean vectors (with 
unstructured positive-definite covariance matrices). These statistics, as \citet{Schott07} says, may be seen as generalizations of the statistic proposed by \citet{BaiSaranadasa96}, in as much the same way as the LRT for the test of equality of several population mean vectors may be seen as a generalization of Hotelling's $T^2$.
However, all these testing procedures need to impose some restrictions on the population covariance matrix, in order that the asymptotic Normal distributions derived hold, a problem that was overcame by \citet{Zhangetal2020} by using an approximate non-normal distribution for their test statistic. 

One other approach that was developed was that of considering tests based on pooled univariate t-tests, eliminating this way the need to use the inverse of what would be, in the high-dimensional setting, non-invertible covariance matrices \citep{Wuetal2006,Gregoryetal2015}.
Yet another path of research transforms the data in trying to use then the existing tests for the non high-dimensional case, by using
random projections, and, in general, projections on a space with a reduced dimension, while trying to keep the original space the least deformed possible. Among authors that took this path we may mention \citet{Lopesetal2011} and \citet{Srivetal2016}.
Ridge regularization of the covariance matrix and LASSO techniques have also been proposed as adequate for MANOVA in the high-dimensional setting \citep{Ullah2015}, which are techniques that may be related to a certain extent to dimensional reduction techniques.

One more approach is the one that uses Bayesian techniques, namely Bayes factors, which are indeed not that far from LRTs. But in its essence this technique may end up suffering from the same problems as common LRTs and the Hotelling $T^2$ statistic suffer from.  However, by combining the Bayes factor technique with a random projection technique on a lower dimensional space, \citet{Zohetal2018}, were able to develop a test for the equality of $q=2$ population mean vectors in high-dimensions.

However we do not intend to use any transformed statistic, neither any data reduction technique nor any Bayesian approach, and even less any pooled univariate tests. Also, we do not want to have to impose any restrictions on the population covariance matrix in order to be able to obtain an asymptotic Normal distribution for the test statistic. Moreover, most procedures developed so far also need the sample sizes to converge to infinity in order that the Normal asymptotic distributions hold, which is indeed quite a incongruity in methods that we want to apply to small sample sizes.
Rather we want to show that it is possible, by reducing the number of parameters associated with the
covariance matrix to be estimated, still allowing this population covariance matrix to have a quite encompassing structure, to successfully apply an LRT for the MANOVA in the high-dimensional setting, actually enjoying a number of significant advantages, namely the existence of an asymptotic distribution which is quite simple to obtain and that in order to hold only needs the number of variables to grow large, with no restrictions at all on the sample sizes.

Admittedly, if one asks authors and researchers in the area of Statistics if it would be possible to develop a likelihood ratio test (LRT) for high-dimensional one-way {\sc manova}, the answer would most likely be a resounding ``No''. However, the right answer is actually ``Yes''! How can it be? By `high-dimensionality' we understand a situation where the overall sample size is less than the number of variables involved. Isn't it true that in the common LRT for one-way {\sc manova} one needs to have an overall sample size ($n$) that is larger than the sum of the number of variables ($p$) plus the number of populations ($q$) involved? Yes, it is. But this is the case because we use covariance matrices that are assumed just to be positive-definite, with no further structure, and that as such, require the estimation of $p(p+1)/2$ parameters. But one can circumvent this problem by assuming some structure for the covariance matrices, which is at the same time wide enough to not `fall too far from the merely positive-definite unstructured structure' and that is able to accommodate as particular cases a number of important covariance structures commonly used and commonly seen as useful, while being at the same time parsimonious enough to allow for a remarkable reduction of the number of parameters, associated with it, to be estimated during the testing process.

It is known that commonly LRT statistics used in Multivariate Analysis have, under the corresponding null hypotheses, a distribution that is the same as that of a product of a number of independent Beta r.v.'s.
As such, the ideal choice for the covariance structure would be one that would make the distribution of the LRT statistic to depend on $p$ only through the number of such r.v.'s involved in its distribution, while allowing the parameters of such r.v.'s to not depend on $p$.
Is such a choice possible to achieve? The answer is ``yes'', as it will be shown in the next section. Actually it is possible to choose this structure in such a way that the LRT performs very well for all types of covariance matrices and for all possible sample sizes, both larger or smaller than the number of variables involved, making of it a very general purpose test. And when we refer to samples that in size are smaller than the number of variables involved, we refer to situations in which the overall sample size, that is,
for a number of $q$ populations being tested, the sum of the sizes of the $q$ samples, has to be only larger than $q$, what only requires that at least one of the samples has to have size 2, with all other samples being allowed to have just size 1.
But, is it possible to use an LRT satisfactorily is such situations? The answer is ``Yes''. In most cases we will even be able to use the exact distribution of its associated statistic in a quite simple form, and for moderately large values of $p$, usually just larger than 8, we will also be able to use a nice asymptotic normal distribution.

And what about the Type I error performance of the proposed test?
Indeed, as \citet{Konetal2021} refer, when dealing with high-dimensional data ``The first statistical problem at hand is neither the high dimensionality of the data nor the relatively low statistical
power of the tests when sample sizes are very small -- it is the accurate type-1 error rate control of the methods''.
It happens that the proposed test, opposite to most existing tests, displays in general an excellent control of the Type I error rate for all other situations, including non-Normal, extremely skewed or leptokurtic, heavy tailed, distributions. Furthermore, given its `origin', it also displays an exact Type I error for underlying multivariate Normal r.v.'s, with the covariance structure stated in Section 2.

The remaining two questions are: (i) how does such a test perform, in the overall, when compared with existing tests developed for the high-dimensional {\sc manova}? (ii) being developed for normally distributed r.v.'s, will it really be applicable to non-normal r.v.'s?

The answer to the first question is: `it will even be able to outperform most existing tests in most situations', and the answer to the second question is: `amazingly enough, and although being originally developed for underlying Normal r.v.'s, it is really applicable to all kinds of distributions, even those that are extremely skewed, and even to continuous Uniform distributions and even to test differences in the location parameters of Cauchy distributions.

The answers to the two questions above can only be given through the implementation of rather extensive and encompassing simulations, what is done in Section 3.

The advantages of the proposed LRT may be summarized as being:
\begin{itemize}[noitemsep,nolistsep]
\item[--] it is easily applicable either to two or more than two populations
\item[--] it has no requirement on the relationship between the sample size and the dimension, that is, the number of variables involved, with no other requirement than ${n>q}$, that is, the overall sample size larger than the number of populations being tested
\item[--] it has both the exact and a sharp, easy to compute, asymptotic Normal distribution available
\item[--] it is rather easy to compute
\item[--] it has an extremely good behavior for extremely small samples
\item[--] it has a remarkably good behavior, both in terms of power as well as in terms of Type I error rate, for a large array of distributions, including highly skewed and leptokurtic distributions
\end{itemize}

The convenience of using such an LRT relies also on the fact that then the results and theory related with LRTs will be enough to establish our results, including those related with the asymptotic distribution of the test statistic, with no need for the derivation of any further elaborate results, leading to a quite simple, but extraordinarily effective approach.

\subsection{The quest for power and the problem of inflated power levels}

To use a test that has higher power is certainly a good goal, but we have to be aware of the risk of using tests that may give inflated power values, and that are not able to adequately control the Type I error levels, since inflated Type I error levels are a problem that may lead one to reject hypothesis that should indeed not be rejected, resulting in the {\sc manova} model in assuming the existence of differences between levels of the factor where they actually do not exist.
Usually these two aspects, inflated Type I error levels and inflated power values, are actually related, due to the fact that, to start with, the asymptotic null distributions used already have some problems in terms of fitting the whole set of values that the test statistic displays under the null hypothesis.

Indeed many tests used for high-dimensional one-way {\sc manova} display, under the null hypothesis, inflated Type I error rates, what leads to deformed power functions, which ultimately lead to inflated power levels.

By conducting extensive simulations, in Section 4, with a very large number of iterations for each case we are able to obtain what may be considered a good sample of the values under the null hypothesis for each of the several statistics used in each case. Their sample quantiles under the null hypothesis may then be used to estimate from the set of values obtained under the alternative hypothesis which should be the power values for given $\alpha$ values. These power values may be seen as the `true' power values that should be displayed by each test in that given situation, and we may compare them with the power values obtained from the exact and asymptotic distributions obtained for these test statistics.

The test statistic we propose has an associated statistic for which both the exact and the asymptotic distribution obtained display a particularly good fit for a large array of situations and distributions, that is, for a large array of possible underlying distributions for the original data, and as such a good control of the Type I error rate and in general non-inflated power values.

\subsection{The test statistics used for comparison with the proposed test}

The tests chosen to compare in Section 4 the proposed LRT with were the tests from \citet{Fujietal04,Schott07,ChenQin10} and \citet{Zhangetal2020}.

The tests from \citet{Fujietal04} and \citet{Schott07} were chosen given the fact that they may be used for more than two populations and also because, as already mention before, they use exactly the within and between covariance matrices used in the common LRT for the test of equality
of mean vectors. One further reason is that, as may be seen from the results in Section 4, we also want to show that these two tests are intrinsically the same test.

The test from \citet{ChenQin10}, although being applicable only to the case ${q=2}$, was chosen since it may seen as representing a slim version of the test from \citet{BaiSaranadasa96}, where, also as already mentioned above, the cross-product terms are not included, and which distribution the authors refer was derived without assuming the equality of the two population covariance matrices, so that we want to check its performance in these situations and compare it with that of the test we propose. Also, we think that actually the removal of the cross-product terms from the test statistic may lead to a less good capability to handle non-normal populations, as it will be shown in Section 4.

The test from \citet{Zhangetal2020} was chosen given the fact that it presents a different way to handle the approximation to the distribution of the test statistic, which in this case is a Gamma distribution, and also given the fact that the authors present an approach that supposedly allows for the estimation of the parameters in the distribution of the test statistic that will work well both for Normal ans well as for non-Normal distributions.

There are a few brief details about the computation of these test statistics that should be mentioned here, in order to make it easier to spot in the above references the expressions that were used in the computations of these tests.

The test statistic from \citet{Fujietal04} was computed using the expression given in the Introduction section of \citet{Schott07} paper, which corresponds indeed to use the original expression for the test statistic from \citet{Fujietal04} in the first expression in subsection 2.1 of their paper and then use for the estimator of the standard deviation an expression similar to the one that appears on page 22 of their paper, but where the $p$'s that appear in both the numerator and the denominator should not be there, since they in the correct expression they should cancel out.

For \citet{Schott07} test we use the formulation of the test statistic $t_{np}$ on page 1827 of the paper and the expression for the estimator of the variance that appears on page 1829, associated with expression (5).

For the test from \citet{ChenQin10} we use the first expression after expression (2.4) on page 811 for the computation of the test statistic and the last expression in section 2 of their paper to compute the estimates of the variance, with the traces replaced by their estimators given on page 814 of the same paper.

For the test from \citet{Zhangetal2020} we used expression (3) in their paper to compute the test statistic and then used the distribution in expression (12) as the one for the approximate distribution of the test statistic, with the estimates of the parameters $d$ and $\beta$ given by expression (24).

\section{The covariance structure chosen and the LRT}

\subsection{Circular or circulant matrices}


Circular or circulant matrices are a particular case of Toeplitz matrices and they have a set of properties which makes them suitable for a wide range of applications \citep{Mukherjee}, which include data compression \citep{Duetal,Carrasquinha}. Indeed, although it is not the case that most of the elements of a circular matrix are null, circular matrices, and namely circular covariance matrices, which are symmetric, may be considered sparse matrices, since a $p\by p$ circular matrix is defined by a set of parameters much smaller than $O(p^2)$, but rather of the order $O(p)$, and as such may be stored in an $O(p)$ number of memory positions. Actually, as already mentioned in the last section, a circular covariance matrix is defined by only $\left\lfloor\frac{p}{2}\right\rfloor+1$ parameters.
Also, circular or circulant covariance matrices appear naturally as covariance matrices of circular processes associated with stationary time series \citep{Andersontimeseries,Fuller,Pollock} and as particular cases of the Mat\'ern covariance function \citep{Matern}, widely used in spatio-temporal studies \citep{Minasny,Cressie,Sherman} and machine learning \citep{Rasmussen},
when the observation points are placed at the vertices of regular polygons.
(See Appendix A for a more detailed, although quite brief, account of some of the properties and applications of circular or circulant matrices.)

As such, the circular covariance structure seems to be a possible good choice in order to fulfill our objective.

Let us assume that
\[
\u\bX_k\sim N_p(\u\bmu_k,\bSigma)\,,~~~~k=1,\dots,q
\]
and that we want to test the null hypothesis
\be
H_0: \u\bmu_1=\dots=\u\bmu_q\,.
\label{hyp}
\ee

We will assume for $\bSigma$ a circular covariance structure, which is a structure that is not `too far away' from the unstructured positive-definite structure (we will elaborate on this a little more later in this section) and that is rich enough to have as particular cases other covariance structures of interest.

We say that a covariance matrix $\bSigma=\left[\sigma_{ij}\right]$ is circular if for
\[
\u\bX_k=\left[X_{k1},\dots,X_{kp}\right]'
\]
we have, for $i=1,\dots,p$ and $\ell=0,\dots,\lfloor p/2\rfloor$,
\[
\sigma_{i,\mod^*(i+\ell,p)}=\sigma_{i,\mod^*(i+p-\ell,p)}=Cov\left(X_{ki},X_{k,\mod^*(i+\ell,p)}\right)=Cov\left(X_{ki},X_{k,\mod^*(i+p-\ell,p)}\right)=\sigma_\ell
\]
where
\[
\mod^*(a,b)=\left\{
\ba{ll}
\mod(a,b) & \hbox{if }a\neq b\ms\\
b & \hbox{if }a=b\,.
\ea
\right.
\]

This entails for $\bSigma$, for example, respectively for $p=6$ and $p=7$ a structure like
\[
\bSigma=\left[
\ba{cccccc}
\sigma_0 & \sigma_1 & \sigma_2 & \sigma_3 & \sigma_2 & \sigma_1 \ms\\
\sigma_1 & \sigma_0 & \sigma_1 & \sigma_2 & \sigma_3 & \sigma_2 \ms\\
\sigma_2 & \sigma_1 & \sigma_0 & \sigma_1 & \sigma_2 & \sigma_3 \ms\\
\sigma_3 & \sigma_2 & \sigma_1 & \sigma_0 & \sigma_1 & \sigma_2 \ms\\
\sigma_2 & \sigma_3 & \sigma_2 & \sigma_1 & \sigma_0 & \sigma_1 \ms\\
\sigma_1 & \sigma_2 & \sigma_3 & \sigma_2 & \sigma_1 & \sigma_0 
\ea
\right]~~~~{\rm and}~~~~
\bSigma=\left[
\ba{ccccccc}
\sigma_0 & \sigma_1 & \sigma_2 & \sigma_3 & \sigma_3 & \sigma_2 & \sigma_1 \ms\\
\sigma_1 & \sigma_0 & \sigma_1 & \sigma_2 & \sigma_3 & \sigma_3 & \sigma_2 \ms\\
\sigma_2 & \sigma_1 & \sigma_0 & \sigma_1 & \sigma_2 & \sigma_3 & \sigma_3 \ms\\
\sigma_3 & \sigma_2 & \sigma_1 & \sigma_0 & \sigma_1 & \sigma_2 & \sigma_3 \ms\\
\sigma_3 & \sigma_3 & \sigma_2 & \sigma_1 & \sigma_0 & \sigma_1 & \sigma_2 \ms\\
\sigma_2 & \sigma_3 & \sigma_3 & \sigma_2 & \sigma_1 & \sigma_0 & \sigma_1 \ms\\
\sigma_1 & \sigma_2 & \sigma_3 & \sigma_3 & \sigma_2 & \sigma_1 & \sigma_0 
\ea
\right]
\]

It is easy to see that the circular covariance structure has as particular cases the equivariance-equicorrelation structure studied by \citet{Wilks46}, which is sometimes also called the compound symmetric covariance structure, and the spherical structure, where all covariances are assumed to be null and all the variances equal.

Let us suppose we have $q$ independent samples, one from each $\u\bX_k$ and that the sample from $\u\bX_k$ has size $n_k$, and let us take $n=\sum_{k=1}^q n_k$. Then the $(2/n)$-th power of the LRT statistic to test the null hypothesis in \nref{hyp}, assuming for $\bSigma$ a circular structure, is \citep{Coe18}
\be
\Lambda=\prod_{j=1}^p \frac{v_j^*}{v_j^{**}}
\label{lambda}
\ee
where
\[
v_j^*=\left\{
\ba{ll}
a_{jj}^*, & j=1\hbox{ and }j=m+1\hbox{ if $p$ is even}\ms\\
(a^*_{jj}+a_{p-j+2,p-j+2}^*)/2, & j=2,\dots,p-m,m+2,\dots,p
\ea
\right.
\]
and
\[
v_j^{**}=\left\{
\ba{ll}
c_{jj}^*, & j=1\hbox{ and }j=m+1\hbox{ if $p$ is even}\ms\\
(c^*_{jj}+c_{p-j+2,p-j+2}^*)/2, & j=2,\dots,p-m,m+2,\dots,p
\ea
\right.
\]
with $a^*_{jj}$ and $c_{jj}^*$ denoting, respectively, the $j$-th diagonal elements of the matrices
\be
\bs A^*=\bs{UAU}'~~~~{\rm and}~~~~\bs C^*=\bs{U(A+B)U}'
\label{mats}
\ee
where $\bs U$ is an orthogonal matrix with elements
\be
u_{ij}=\frac{1}{\sqrt{p}}\left\{\cos\left(2\pi(i-1)(j-1)/p\right)+\sin\left(2\pi(i-1)(j-1)/p\right)\right\}\,,~~~~i,j\in\{1,\dots,p\}
\label{runningu}
\ee
and
\[
\bs A=\sum_{k=1}^q (n_k-1)\bs S_k~~~~{\rm and}~~~~\bs B=\sum_{k=1}^q n_k\left(\ov{\u\bX}_k-\ov{\u\bX}\right)\left(\ov{\u\bX}_k-\ov{\u\bX}\right)'
\]
are, respectively, the `within' and `between' sum of squares and sums of products matrices, with $\bs S_k$ and $\ov{\u\bX}_k$ denoting respectively the sample covariance and mean vector of the sample from $\u\bX_k$, and where
\[
\ov{\u\bX}=\frac{1}{n}\sum_{k=1}^q n_k\ov{\u\bX}_k
\]
is the overall mean vector.

Then, it is possible to show that under the null hypothesis in \nref{hyp},
\be
\Lambda\st Y_1(Y_2)^{\mod(p+1,2)}\prod_{j=2}^{p-m} (Y_j^*)^2
\label{dist1}
\ee
where `\,$\st$' means `stochastically distributed as', and
\be
Y_1\equivd Y_2\sim Beta\left(\frac{n-q}{2},\frac{q-1}{2}\right)~~~~{\rm and}~~~~Y_j^*\sim Beta(n-q,q-1)~~(j=2,\dots,p-m)
\label{dist2}
\ee
are a set of $p-m-1+\mod(p+1,2)$ independent random variables, and where `\,$\equivd$' means `is equivalent in distribution to' \citep{Coe18}.

But then, from \nref{dist2} we may see that the distribution of $\Lambda$ holds as long as $n>q$, what shows that we can carry out the test as long as ${q-1}$ of the $q$ samples have size just 1, and the remaining one with size at least 2, making thus this test perfectly adequate to be used in a high-dimensional setting.

We should note that for the generality of the existing tests for high-dimensional {\sc manova} the use of samples of size 1 for any of the $\u\bX_k$ is not viable, but for the present test presents absolutely no problem, as long as at least one of the samples has at least size 2.
From all the `logic' and intuition one can place in the realm of a test for one-way {\sc manova} this is indeed the limit situation for the possible implementation of a test in that setting, being indeed the case that most of the test developed for high-dimensional {\sc manova} do not support such situations.

For odd $q$ the statistic $\Lambda$ in \nref{lambda} has an exact distribution which is very manageable and which assumes the form of an EGIG (Exponentiated Generalized Integer Gamma) distribution \citep{Arnetal2013}, with probability density function (pdf) and cumulative distribution function (cdf) respectively given by \citep{Coe18}
\be
f_\Lambda(z)=f^{\hbox{\tiny EGIG}}\left(z\,\Bigl|\,\{r_\ell\}_{\ell=1:q-1},\left\{\frac{n-q+\ell-1}{2}\right\}_{\ell=1;q-1};q-1\right)
\label{pdfoddq}
\ee
and
\be
F_\Lambda(z)=F^{\hbox{\tiny EGIG}}\left(z\,\Bigl|\,\{r_\ell\}_{\ell=1:q-1},\left\{\frac{n-q+\ell-1}{2}\right\}_{\ell=1;q-1};q-1\right)
\label{cdfoddq}
\ee
with ${0<z<1}$ and
\be
r_\ell=\left\{
\ba{ll}
\lfloor p/2\rfloor+1\,, & \hbox{odd }\ell\ms\\
p-\lfloor p/2\rfloor-1\,, & \hbox{even }\ell
\ea
\right.
\label{rell}
\ee
where the notation used is the one in \citet[App.\,B]{Arnetal2013}.
For even $q$ the exact distribution of $\Lambda$ is no more easily manageable but it is possible to obtain near-exact distributions as finite mixtures of GNIG (Generalized Near-Integer Gamma) distributions, with pdf's and cdf's of the form \citep{Coe18}
\be
f_\Lambda(z)=\sum_{k=0}^{m^*} f^{\hbox{\tiny GNIG}}\left(\log\,z\,\Bigl|\,\{r_\ell\}_{\ell=1:q-1},r+k;\left\{\frac{n-q+\ell-1}{2}\right\}_{\ell=1:q-1},\lambda^*;q\right)\frac{1}{z}
\label{pdfevenq}
\ee
and
\be
F_\Lambda(z)=\sum_{k=0}^{m^*}\left(1-F^{\hbox{\tiny GNIG}}\left(\log\,z\,\Bigl|\,\{r_\ell\}_{\ell=1:q-1},r+k;\left\{\frac{n-q+\ell-1}{2}\right\}_{\ell=1:q-1},\lambda^*;q\right)\right)
\label{cdfevenq}
\ee
where
\[
r=\left\{
\ba{ll}
1\,, & \hbox{even }p\ms\\
1/2\,, & \hbox{odd }p\,,
\ea
\right.
~~~~
r_\ell=\left\{
\ba{ll}
\lfloor p/2\rfloor+1\,, & \hbox{odd }\ell, ~\ell\neq q-1\ms\\
p-\lfloor p/2\rfloor -1\,, & \hbox{even }\ell\hbox{ and }\ell=q-1\,,
\ea
\right.
\]
$\lambda^*$ is obtained through the numerical procedure described in \citet{Coe18}, and
$m^*$ is the number of exact moments of $W=-\log\Lambda$ that are matched by the near-exact distribution,
and where the notation used is that in \citet[App.\,1]{CoeMar12}.
We should note that for even $p$, given that $r$ is then an integer, the above near-exact distributions actually turn into mixtures of GIG distributions, and also that there is a small, but important to correct, typo in the expressions for the near-exact pdf and cdf in \citet{Coe18}, where $r$ is to be replaced by $r+k$.

An alternative in order to handle the distribution of $\Lambda$ is to use a numerical inversion module, as the `FT' module programmed in Mathematica and which uses the fixed Talbot method for numerical inversion of Laplace transforms \citep{AbateValko2004}, and which may be used on the characteristic function (CF) of $W=-\log\Lambda$ allowing us to obtain values for the PDF and CDF of $\Lambda$. Although module `FT' is programmed for the inversion of Laplace transforms, it may readily be applicable to the inversion of Fourier transforms, and as such, to the inversion of the CF of $W=-\log\Lambda$, but it only works for random variables in $\mathbb R^+$. Used together with a simple bisection method allows for the quick computation of exact quantiles for $\Lambda$. 

This LRT for the test of equality of mean vectors, with circular or circulant covariance matrices, leads however to only slight gains in power. That is, if the population covariance matrices are indeed circular and we use the usual LRT for unstructured covariance matrices, we will be loosing some power, while the use of the LRT just presented should lead to some gains in power. Unpublished results obtained by the author show that this gain in power is however deceiving small. While this fact may be seen as a drawback for the LRT being presented, it is indeed its great virtue in the setting we are willing to implement it, since this only slight gain in power shows that intrinsically the two LRTs, that is the one for circular covariance matrices and the one for unstructured covariance matrices, are not so much different, and that, as such, the LRT introduced in this paper has good chances of being applied to any type of covariance matrices, probably with some slight bias under the null hypothesis, when the covariance matrices are not circular but only positive-definite.

The question now is: how will the test perform if the covariance matrices of the $\u\bX_k$ are not circular? And one other important question is: since the distribution given by \nref{dist1}-\nref{dist2} was obtained under the normality assumption for the random vectors $\u\bX_k$, how will the test perform for non-Normal random variables, namely for skewed or heavy tailed distributions?
(Of course when using other distributions that not the Normal, the test will not be anymore an LRT!)

Related with these two questions is also the question: and how will its performance compare with other tests developed for the high-dimensional case?

These questions are all related with two other questions, which are: how well does the null distribution derived above fit the test statistic values, under the null hypothesis, for non-normal distributions, and as such, how well does this test control the Type I error rate? And does it show inflated power values when under the alternative hypothesis?

As we will see, the answer to the two above questions is that the null distribution obtained above for the test statistic still holds quite well under a very large array of distributions for the $\u\bX_k$ vectors, even for skewed and heavy tailed ones, with the test statistic displaying thus a very good control on the Type I error rates, while the answer to the second one is `no!', that is, for most situations and most underlying distributions for the $\u\bX_k$, the test does not show inflated power values under the alternative hypothesis.

Due to the fact that in order to give an answer to the above questions we will need to compute values of the test statistic, both under the null as well as under the alternative hypotheses, for a large array of different underlying distributions for the $\u\bX_k$, such questions
may only be answered by resorting to Monte Carlo simulations. This will be done in the next section, after we obtain a Normal asymptotic distribution for $W=-\log\Lambda$, which will be asymptotic for increasing values of $p$, and which will be shown to have a very good behavior even for very small sample sizes and not necessarily so large values of $p$, for a large set of underlying distributions for the $\u\bX_k$.


\subsection{An asymptotic Normal distribution for $\Lambda$}

In this section we show that it is possible to obtain a quite simple asymptotic distribution for the statistic $\Lambda$ in \nref{lambda} and \nref{dist1}, which has no requirements on the sample size, since it is indeed asymptotic for $p$, the number of variables involved, rather than being asymptotic for the sample size. This is actually an important feature of this asymptotic distribution, since we want to apply the test in situations where the sample sizes are small, or even extremely small and the number of variables is large or even very large.

\begin{theorem}
Let, for $m=\lfloor p/2\rfloor$, 
\[
\Lambda\st Y_1(Y_2)^{\mod(p+1,2)}\prod_{j=2}^{p-m} (Y_j^*)^2
\]
where `\,$\st$' means `stochastically distributed as', and
\[
Y_1\equivd Y_2\sim Beta\left(\frac{n-q}{2},\frac{q-1}{2}\right)~~~~{\rm and}~~~~Y_j^*\sim Beta(n-q,q-1)~~(j=2,\dots,p-m)
\]
are a set of $p-m-1+\mod(p+1,2)$ independent random variables, and where `\,$\equivd$' means `is equivalent in distribution to'.

Then
\[
\frac{-\log\Lambda-\left\{(1+\mod(p+1,2)\mu+2(p-m-1)\mu^*\right\}}{\sqrt{(1+\mod(p+1,2))\sigma^2+4(p-m-1)(\sigma^*)^2}}\xrightarrow[p\rightarrow\infty]{d} N(0,1)\,,
\]
for
\be
\mu=E(W_j)=\Psi\left(\frac{n-1}{2}\right)-\Psi\left(\frac{n-q}{2}\right)~~~~{\rm and}~~~~\mu^*=E(W^*_j)=\Psi(n-1)-\Psi(n-q)\,,
\label{mus}
\ee
where $\Psi$ is the digamma function, $W_j=-\log Y_j$ and $W^*_j=-\log Y_j^*$, and
\be
\sigma^2\!=\!Var(W_j)\!=\!\Psi'\left(\frac{n-q}{2}\right)-\Psi'\left(\frac{n-1}{2}\right)~~~~{\rm and}~~~~(\sigma^*)^2\!=\!Var(W^*_j)\!=\!\Psi'(n-q)-\Psi'(n-1)
\label{sigmas}
\ee
where $\Psi'$ is the trigamma function.
\end{theorem}

\begin{proof}
Let $Y_j\sim Beta\left((n-q)/2,(q-1)/2\right)$ and $Y^*_j\sim Beta\left(n-q,q-1\right)$ $(j=1,\dots,p-1)$ be two sets of $p-1$ independent random variables and let $W_j=-\log Y_j$ and $W^*_j=-\log Y_j^*$. Then, for $\mu$ and $\mu^*$ in \nref{mus},
let
\[
\mu_3=E\left(|W_j-\mu|^3\right)\,,~~~~\mu_3^*=E\left(|W^*_j-\mu^*|^3\right)
\]
and consider $\sigma^2$ and $(\sigma^*)^2$ in \nref{sigmas}.

Then we have
\[
\mu_3>\mu_3^*~~~~{\rm and}~~~~\sigma^2>(\sigma^*)^2
\]
and thus
\[
\ba{l}
\ds 0\leq\lim_{p\rightarrow\infty}\frac{(1+\mod(p+1,2))\mu_3+2(p-m-1)\mu_3^*}{\left((1+\mod(p+1,2)\sigma^2+4(p-m-1)(\sigma^*)^2\right)^{3/2}}\ms\\
\hskip 5cm \ds \leq\lim_{p\rightarrow\infty}\frac{\left(2(p-m-1)+1+\mod(p+1,2)\right)\mu_3}{\left(\left(4(p-m-1)+1+\mod(p+1,2)\right)(\sigma^*)^2\right)^{3/2}}=0\,,
\ea
\]
so that for
\[
W=-\log\Lambda\st \left\{
\ba{ll}
W_1+W_2+2\sum_{j=2}^{p-m} W_j^* & \hbox{for even }p\ms\\
W_1+2\sum_{j=2}^{p-m} W_j^* & \hbox{for odd }p
\ea
\right.
\]
Lyapunov's condition \citep[Chap.\,5, Sect.\,27]{Billingsley} holds for $\delta=1$, and thus the result follows.
\end{proof}

We may thus say that, for increasing $p$,
\[
\ba{l}
\ds W=-\log\Lambda\as N\left(1+\mod(p+1,2)\mu+2(p-m-1)\mu^*,\right.\ms\\
\hskip 6.5cm \ds \left. 1+\mod(p+1,2))\sigma^2+4(p-m-1)(\sigma^*)^2\right)\,.
\ea
\]

As we will see, from the results of the simulations in Sect.\,\ref{Simul}, this asymptotic distribution closely holds even for quite small $p$ as well as for a wide range of underlying distributions for the $\u\bX_k$, that are not necessarily either Normal or symmetric.

In terms of the rate of convergence, this asymptotic distribution has a convergence rate of $1/\sqrt{p-1}$, since it is based on a CLT for a sum of ${p-1}$ independent random variables. This means that, for example, if we use the measure 
\[
\Delta=\frac{1}{2\pi}\int_{-\infty}^{+\infty}\left|\frac{\Phi_W(t)-\Phi(t)}{t}\right|dt\,,
\]
where 
\[
\Phi_W(t)=\left(\frac{\Gamma\left(\frac{n-1}{2}\right)\Gamma\left(\frac{n-q}{2}-\ii t\right)}{\Gamma\left(\frac{n-q}{2}\right)\Gamma\left(\frac{n-1}{2}-\ii t\right)}\right)^{\!\!1+\mod(p+1.2)}\,\prod_{j=2}^{p-m}\frac{\Gamma(n-1)\,\Gamma(n-q-2\ii t)}{\Gamma(n-q)\,\Gamma(n-q-2\ii t)}
\]
is the characteristic function (c.f.) of $W=-\log\Lambda$, and $\Phi(t)$ is the c.f.\ of the $N(0,1)$ distribution, is a tight upper bound on the difference between the exact c.d.f.\ of $W$ and the $N(0,1)$ c.d.f., with
\[
\Delta\geq\max_{w>0}\left|F_W(w)-F(w)\right|,
\]
where $F_W(\,\cdot\,)$ is the exact c.d.f.\ of $W$ and $F(\,\cdot\,)$ is the c.d.f.\ of the $N(0,1)$ distribution.

The rate of convergence in Theorem 1 being equal to $1/\sqrt{p-1}$ means that if for a given $q$, a given value of $n$ and a given value of $p=p_0$, the measure $\Delta$ has the value $\Delta_0$, then for $p^*=kp_0$, the measure $\Delta$ will have a value approximately equal to 
\[
\Delta_0\Bigr/\sqrt{\frac{p_0-1}{p^*-1}}=\Delta_0\Bigr/\sqrt{\frac{p_0-1}{kp_0-1}}\,.
\]

\section{Random sample sizes}

Quite often, if not most often, sample sizes are actually not set up `a priori' by the researcher but rather defined by a set of random events, opportunities and constraints. as such, they should be taken as random rather than as fixed.

However, given the way the distributions, which are commonly asymptotic distributions, are obtained for the testing procedures used for high-dimensionality, it is usually not possible to even think of implementing this approach for the tests commonly developed. This is the case mostly because although being developed for high-dimensional cases, and as such, for situations where the sample sizes may be quite small, the asymptotic distributions they use for the test statistics assume this sample size converging to infinity, together with the number of variables.

But, given the characteristic of the LRT proposed in this paper, there is no problem in implementing the random sample size approach, rendering it indeed quite simple, not only for the exact or near-exact distributions, but also for the asymptotic distribution.

The only thing one has to do is to consider now the distribution of the LRT statistic as a mixture of the previously stated distributions, weighted by the probabilitiesof a given sample size. More precisely, let $N$ be now the random variable that represents the overall sample size, from which now $n$ represents a possible value, and let $P(N=n)$ represent the, for a given assumed distribution for $N$, the probability that the r.v.\ $N$ assumes the value $n$. Then the p.d.f.\ of the distribution of $\Lambda$ in \nref{lambda} for the case of odd $q$, will be, from \nref{pdfoddq} and \nref{cdfoddq}, respectively given by
\be
f_\Lambda(z)=\sum_{n=q+1}^u P(N=n)\,f^{\hbox{\tiny EGIG}}\left(z\,\Bigl|\,\{r_\ell\}_{\ell=1:q-1},\left\{\frac{n-q+\ell-1}{2}\right\}_{\ell=1;q-1};q-1\right)
\label{pdfoddqrss}
\ee
and
\be
F_\Lambda(z)=\sum_{n=q+1}^u P(N=n)\,F^{\hbox{\tiny EGIG}}\left(z\,\Bigl|\,\{r_\ell\}_{\ell=1:q-1},\left\{\frac{n-q+\ell-1}{2}\right\}_{\ell=1;q-1};q-1\right)
\label{cdfoddqrss}
\ee
where $r_\ell$ is given by \nref{rell}, and $u$ is the upper bound of the set of possible values of the r.v.\ $N$. In some cases $u$ will have a finite value, as it is the case if we assume for $N$ a Binomial distribution, or it may be infinity, as it is the case if we assume for $N$ a Poisson or a Negative Binomial distribution.

For the case of even $q$, from \nref{pdfevenq} and \nref{cdfevenq}, we will have the near-exact distribution of $\Lambda$ with a p.d.f.\ and c.d.f.\ respectively given by
\be
f_\Lambda(z)=\!\!\sum_{n=q+1}^u\!\! P(N\!=\!n)\,\sum_{k=0}^{m^*} f^{\hbox{\tiny GNIG}}\!\left(\log\,z\,\Bigl|\,\{r_\ell\}_{\ell=1:q-1},r+k;\left\{\frac{n-q+\ell-1}{2}\right\}_{\ell=1:q-1},\lambda^*;q\right)\frac{1}{z}
\label{pdfevenqrss}
\ee
and
\be
F_\Lambda(z)=\!\!\sum_{n=q+1}^u\!\! P(N\!=\!n)\,\sum_{k=0}^{m^*}\left(1-F^{\hbox{\tiny GNIG}}\!\left(\log\,z\,\Bigl|\,\{r_\ell\}_{\ell=1:q-1},r+k;\left\{\frac{n-q+\ell-1}{2}\right\}_{\ell=1:q-1},\lambda^*;q\right)\right),
\label{cdfevenqrss}
\ee
while for the asymptotic distribution of $\Lambda$ we will say that it is, in this case, a mixture of the Normal distributions in Theorem 1, with weights $P(N=n)$, that is, the asymptotic p.d.f.\ of $\Lambda$ will be given by
\be
f^*_\Lambda(z)=\sum_{n=q+1}^u P(N=n)\,\frac{1}{\sqrt{2\pi\sigma^2}}\,e^{-\frac{(z-\mu)^2}{2\sigma^2}}
\label{pdfasymprss}
\ee
for
\[
\mu=1+\mod(p+1,2)\mu+2(p-m-1)\mu^*
\]
and
\[
\sigma^2=1+\mod(p+1,2))\sigma^2+4(p-m-1)(\sigma^*)^2\,,
\]
for $\mu$, $\mu^*$, $\sigma^2$ and $\sigma_2^*$ defined as in the statement of Theorem 1.

Let us now denote by $N_k$ $(k=1,\dots,q)$ the r.v.'s that represent the sample sizes for each of the $q$ populations, with $N=\sum_{k=1}^q N_k$.

We may assume for the $N_k$, and consequently for $n$, a number of different distributions, according to the situation. By doing this we are indeed giving the test procedure a great flexibility, but, as it always happens in Statistics, we will have to pay this flexibility by requiring from the researcher a quite good knowledge of the populations being sampled, in order to be able to set up the correct parameters for the distributions of the $N_k$.

For example, if we are dealing with samples from individuals affected by a given disease or situation and we know that the $k$-th population has say a total of $N^*_k$ individuals, with a prevalence of that disease or situation of $p_k$, then, in case $N_k^*$ is very large and $p_k$ quite small, then we may think about assuming for $N_k$ a Poisson distribution with parameter $\lambda_k=N_k^*p_k$, while if $N_k^*$ is not that large, and/or $p_k$ is not that small, 
we may then think about assuming for $N_k$ a Binomial$(N_k^*,p_k)$ distribution. Then, $N$ would have a Poisson$(\lambda)$ distribution, with $\lambda=\sum_{k=1}^q \lambda_k$, in the first case, and, in the second case, if we assume all the $p_k$ equal to say $p$, a Binomial$(N^*,p)$ distribution, with $N^*=\sum_{k=1}^q N^*_k$.

A Negative Binomial sample size may be assumed for the $k$-th population if we assume that we are sampling `individuals' till a given number $r_k$ of successes or failures is attained. In case we adopt for each of the $q$ populations the same value for the probability of success, say $p$, then $N$ will have a Negative Binomial distribution with parameters $r^*=\sum_{k=1}^q r_k$, and $p$.

Although the sampling schemes related with the Binomial and Negative Binomial distributions for the sample sizes formally should assume a sampling done with replacement, if the values of the population size in each population is rather large, the use of the Binomial or Hipergeometric distributions will lead to very similar results, with a gain in manageability when using the Binomial distributions, given that once the Binomial or Negative Binomial distributions are assumed for the sample size of each population, with the restriction of equal probabilities of success for all $q$ populations, this will lead also to binomial or Negative Binomial distributions for the overall sample size $N$.

As we see from expressions \nref{pdfoddqrss}, \nref{cdfoddqrss}, \nref{pdfevenqrss}, \nref{cdfevenqrss} and \nref{pdfasymprss}, it is only the distribution of $N$, and not that of the $N_k$, that mwtters for the distribution of $\Lambda$, and in computing $P(N=n)$ we have to take into account that only values of $n\geq q+1$ are taken into account, so that for a given $q$, the values $P(N=n)$ have to be taken as
\[
P(N=n)=f_N(n)\Bigl/\left(1-\sum_{i=0}^q f_N(i)\right)
\]
for $n=q+1,\dots,u$, where $f_N(n)$ represents the value of the probability mass function that is being used for $N$ (Poisson, Binomial or Negative Binomial) at the given value $n$, so that in this way we have
\[
\sum_{n=q+1}^u P(N=n)=1\,.
\]

\section{Computation modules for the distributions and the simulations}

All computation modules were programmed in Mathematica, and our choice was to produce a Mathematica notebook with all modules. In this notebook, before any computations are carried out one has to run the first cell that has the modules and then also run the following cell that has the commands that will enable to use the necessary R functions that are used to generate some of the distributions used in the simulations. Details about these distributions are given in the next section. A more detailed description of the modules provided and their use is available in the supplementary material of this paper.

In a summary, there are modules available to compute the exact PDF and CDF of the LRT statistic $\Lambda$ in \nref{lambda} for the cases where $q$ is odd, and to compute the near-exact PDF and CDF for the cases where $q$ is even. There are also modules to compute the quantiles, both by inversion of the CF, as well as by using either the exact or the near-exact CDFs, according to the case. The module that computes the quantiles by inversion of the CF uses the expression of the CF in \citet{Coe18} and the module `FT' from \citet{AbateValko2004}. For moderately large values of $p$ and for odd $q$ the module for the computation of quantiles that uses the exact CDF is quite a bit faster than the module that uses the inversion technique, while for even $q$ the module that uses the inversion of the CF is quite a bit faster than the module that uses the near-exact CDF.

\section{Simulation studies}
\label{Simul}

In this section we are going to apply the test in Section 2 to a wide array of distributions, that is, we will use several different pseudo-random generated values for a number of different multivariate distributions for the random vectors $\u\bX_k$ and evaluate the performance of this test. We will also consider the test from \citet{Fujietal04}, \citet{Schott07} and \citet{ChenQin10} to compare their performances with the LRT introduced in Section 2.

The multivariate distributions used are:
\begin{itemize}
\item[1)] the multivariate Normal, with
\begin{itemize}
\item[a)] a positive-definite covariance matrix that is not circular
\item[b)] a circular covariance matrix
\item[c)] a compound symmetric, or rather, a equivariance-equicovariance matrix
\item[d)] a diagonal matrix, with different diagonal elements
\item[e)] a spherical cobariance matrix
\end{itemize}
\item[2)] a multivariate $T$ distribution with 5, 3 and 1 degrees of freedom, and the same covariance matrices as the ones used for the multivariate Normal distribution
\item[3)] a Dirichlet distribution
\item[4)] a multivariate Uniform distribution with all individual random variables independent
\item[5)] The multivariate Skew-Normal distribution from Azallini
\item[6)] The multivariate Skew-T distribution from Azallini
\item[7)] a multivariate Cook-Johnson's Uniform distribution, with non-independent components
\item[8)] a multivariate generalized Lomax distribution
\item[9)] a multivariate Lomax distribution
\item[10)] a multivariate Burr distribution
\item[11)] a multivariate Mardia-Pareto distribution
\item[12)] a multivariate $F$ distribution
\item[13)] a multivariate inverted Beta distribution
\item[14)] a multivariate Logistic distribution.
\end{itemize}

One of the objectives of using the tests from \citet{Fujietal04} and \citet{Schott07} is also to show that actually these two tests are mostly the same test. While their formulations are somewhat close, it is not an easy task to show that analytically that they are equivalent. However, through the simulations carried out we may see that it is really a fact.

The tests of \citet{ChenQin10} and \citet{Zhangetal2020} are only applicable to two populations, that is, to the case ${q=2}$. Anyway we thought it was important to compare the proposed LRT with a test specifically developed for ${q=2}$ populations. This entailed the need to extend the simulations, for cases of both ${q=2}$ and ${q>2}$. For the case ${q>2}$, and to not make the manuscript even longer than what it already is, we decided to include only the case ${q=4}$. For each case, and both for the cases under the null hypothesis and the alternative hypothesis, 100,000 values of each statistic were computed.

All computations were carried out on an HP computer with a i5 processor, running the Windows 10 operating system and all computations were done with the software Mathematica (version 11.2).
While the pseudo-random values generation for the distributions in 1)--4) above is directly available in Mathematica, for the distributions in 5)--6)
the R package `SN' \citep{Azzalini} was used, while for the distributions in 7)--12) the R package `NonNorMvtDist'
\citep{LunKhat} was used for the pseudo-random values generation, from within Mathematica, after implementing these two packages in workspaces of version 3.3.0 of R.

We should note that the multivariate $T$ distribution with 1 degree of freedom, mentioned in 2) above, is really a multivariate Cauchy distribution, and that, as such, the random variables involved have no expected value and no variance, and as such the test being carried out may only, at best, be seen as a test for location. Anyway, as we will see, all four tests perform yet rather well with this distribution, although with a quite clear advantage for the LRT.



\section{A test for outliers in the high-dimensional case}

At first sight it may seem a bit strange to put together an LRT for {\sc manova} and an outlier test in the same publication, but it happens that actually by using the LRT test developed for the high-dimensional {\sc manova} we can get a very much functional outlier test, which, as the LRT test for {sc manova} can be used both in the high-dimensional case as well as in the non-high-dimensional case.

\section{Conclusions and Future Directions}

Actually, the use of the proposed LRT provides a rather simple approach, yet a very efficient one, with a remarkably good control of the type I error rate for a wide range of distributions, even non-normal, heavy-tailed and highly skewed distributions, and with higher power values for a number of alternative hypotheses situations, namely those in which the random variables involved show moderate to high correlations, yet providing for a quite simple asymptotic distribution of the test statistic for increasing numbers of variables involved, while posing absolutely no restrictions on sampe sizes, which may be as small as just 1 for all the populations involved, with the only requirement that at least one of such samples has to have size at least 2, which are smaller sample sizes requirements than any of the other tests available, being the case that none of them allows for samples as small as 1.



\clearpage

\pagebreak

\section*{Appendix}

\begin{table}[ht!]
\centering
\caption{\label{table1n}Normal distribution with circular covariance matrix with correlations 0.108 to 0.216 and $n_k=4,4$\\[-9pt]}

\end{table}

\vspace{-.5cm}

\begin{table}[ht!]
\centering
\caption{\label{table2n}Normal distribution with circular covariance matrix with correlations 0.108 to 0.216 and $n_k=4,4$\\[-9pt]}
\end{table}

\vspace{-.5cm}

\begin{table}[ht!]
\centering
\caption{\label{table3n}Normal distribution with circular covariance matrix with correlations 0.108 to 0.216 and $n_k=4,4$\\[-9pt]}
\begin{tabular}{ccccccc}
\multicolumn{7}{c}{$\boldsymbol H_0$}\\
 & \multicolumn{2}{c}{LRT} & & & \\[-1pt]
\cline{2-3}\\[-19pt]
$\alpha$ & exact & asymp & \raisebox{5pt}{Fuji} & \raisebox{5pt}{Schott} & \raisebox{5pt}{ChenQin} & \raisebox{5pt}{Zhang}\\
\hline
0.01 & 0.011 & 0.020 & 0.053 & 0.050 & 0.057 & 0.055 \\
0.05 & 0.051 & 0.062 & 0.097 & 0.094 & 0.105 & 0.109\\[7pt]
\end{tabular}
\begin{tabular}{@{\extracolsep{-2pt}}ccccccccccccc}
\multicolumn{12}{c}{~~~~~~~~~~~~~~~~$\boldsymbol H_1$ (shift=+0.5,+0.5)} & \!\!\!\!\!\!\!\\
 & \multicolumn{3}{c}{LRT} & \multicolumn{2}{c}{Fuji} & \multicolumn{2}{c}{Schott} & \multicolumn{2}{c}{ChenQin} & \multicolumn{2}{c}{Zhang} & \!\!\!\!\!\!\!\\[-11pt]
& \multicolumn{3}{c}{\rule{105pt}{.5pt}} & \multicolumn{2}{c}{\rule{69pt}{.5pt}} & \multicolumn{2}{c}{\rule{69pt}{.5pt}} & \multicolumn{2}{c}{\rule{69pt}{.5pt}} & \multicolumn{2}{c}{\rule{69pt}{.5pt}} & \!\!\!\!\!\!\!\\[-4pt]
$\alpha$ & exact & asymp & simul & asymp & simul & asymp & simul & asymp & simul & asymp & simul & \!\!\!\!\!\!\!\\
\hline\\[-8pt]
0.01 & 0.029 & 0.051 & 0.027 & 0.094 & 0.021 & 0.090 & 0.021 & 0.086 & 0.019 & 0.095 & 0.021 \\
0.05 & 0.111 & 0.132 & 0.110 & 0.161 & 0.089 & 0.156 & 0.089 & 0.147 & 0.076 & 0.176 & 0.094
\end{tabular}
\end{table}

\vspace{-.5cm}

\begin{table}[ht!]
\centering
\caption{\label{table4n}Normal distribution with circular covariance matrix with correlations 0.108 to 0.216 and $n_k=4,4$\\[-9pt]}
\begin{tabular}{ccccccc}
\multicolumn{7}{c}{$\boldsymbol H_0$}\\
 & \multicolumn{2}{c}{LRT} & & & \\[-1pt]
\cline{2-3}\\[-19pt]
$\alpha$ & exact & asymp & \raisebox{5pt}{Fuji} & \raisebox{5pt}{Schott} & \raisebox{5pt}{ChenQin} & \raisebox{5pt}{Zhang}\\
\hline
0.01 & 0.010 & 0.020 & 0.052 & 0.050 & 0.056 & 0.055 \\
0.05 & 0.050 & 0.060 & 0.096 & 0.093 & 0.102 & 0.108\\[7pt]
\end{tabular}
\begin{tabular}{@{\extracolsep{-2pt}}ccccccccccccc}
\multicolumn{12}{c}{~~~~~~~~~~~~~~~~$\boldsymbol H_1$ (shift=+0.5,--0.5)} & \!\!\!\!\!\!\!\\
 & \multicolumn{3}{c}{LRT} & \multicolumn{2}{c}{Fuji} & \multicolumn{2}{c}{Schott} & \multicolumn{2}{c}{ChenQin} & \multicolumn{2}{c}{Zhang} & \!\!\!\!\!\!\!\\[-11pt]
& \multicolumn{3}{c}{\rule{105pt}{.5pt}} & \multicolumn{2}{c}{\rule{69pt}{.5pt}} & \multicolumn{2}{c}{\rule{69pt}{.5pt}} & \multicolumn{2}{c}{\rule{69pt}{.5pt}} & \multicolumn{2}{c}{\rule{69pt}{.5pt}} & \!\!\!\!\!\!\!\\[-4pt]
$\alpha$ & exact & asymp & simul & asymp & simul & asymp & simul & asymp & simul & asymp & simul & \!\!\!\!\!\!\!\\
\hline\\[-8pt]
0.01 & 0.036 & 0.063 & 0.035 & 0.086 & 0.016 & 0.082 & 0.016 & 0.077 & 0.013 & 0.088 & 0.016 \\
0.05 & 0.133 & 0.155 & 0.134 & 0.155 & 0.082 & 0.150 & 0.082 & 0.138 & 0.068 & 0.171 & 0.082
\end{tabular}
\end{table}

\vspace{-.5cm}

\begin{table}[ht!]
\centering
\caption{\label{table5n}Normal distribution with circular covariance matrix with correlations 0.108 to 0.216 and $n_k=4,8$\\[-9pt]}
\begin{tabular}{ccccccc}
\multicolumn{7}{c}{$\boldsymbol H_0$}\\
 & \multicolumn{2}{c}{LRT} & & & \\[-1pt]
\cline{2-3}\\[-19pt]
$\alpha$ & exact & asymp & \raisebox{5pt}{Fuji} & \raisebox{5pt}{Schott} & \raisebox{5pt}{ChenQin} & \raisebox{5pt}{Zhang}\\
\hline
0.01 & 0.010 & 0.020 & 0.044 & 0.043 & 0.013 & 0.036 \\
0.05 & 0.051 & 0.062 & 0.085 & 0.084 & 0.036 & 0.084\\[7pt]
\end{tabular}
\begin{tabular}{@{\extracolsep{-2pt}}ccccccccccccc}
\multicolumn{12}{c}{~~~~~~~~~~~~~~~~$\boldsymbol H_1$ (shift=+0.75,+0.75)} & \!\!\!\!\!\!\!\\
 & \multicolumn{3}{c}{LRT} & \multicolumn{2}{c}{Fuji} & \multicolumn{2}{c}{Schott} & \multicolumn{2}{c}{ChenQin} & \multicolumn{2}{c}{Zhang} & \!\!\!\!\!\!\!\\[-11pt]
& \multicolumn{3}{c}{\rule{105pt}{.5pt}} & \multicolumn{2}{c}{\rule{69pt}{.5pt}} & \multicolumn{2}{c}{\rule{69pt}{.5pt}} & \multicolumn{2}{c}{\rule{69pt}{.5pt}} & \multicolumn{2}{c}{\rule{69pt}{.5pt}} & \!\!\!\!\!\!\!\\[-4pt]
$\alpha$ & exact & asymp & simul & asymp & simul & asymp & simul & asymp & simul & asymp & simul & \!\!\!\!\!\!\!\\
\hline\\[-8pt]
0.01 & 0.123 & 0.185 & 0.121 & 0.190 & 0.058 & 0.187 & 0.058 & 0.074 & 0.059 & 0.159 & 0.065 \\
0.05 & 0.316 & 0.353 & 0.314 & 0.307 & 0.209 & 0.304 & 0.209 & 0.156 & 0.199 & 0.301 & 0.229
\end{tabular}
\end{table}

\vspace{-.5cm}

\begin{table}[ht!]
\centering
\caption{\label{table6n}Normal distribution with circular covariance matrix with correlations 0.108 to 0.216 and $n_k=4,8$\\[-9pt]}
\begin{tabular}{ccccccc}
\multicolumn{7}{c}{$\boldsymbol H_0$}\\
 & \multicolumn{2}{c}{LRT} & & & \\[-1pt]
\cline{2-3}\\[-19pt]
$\alpha$ & exact & asymp & \raisebox{5pt}{Fuji} & \raisebox{5pt}{Schott} & \raisebox{5pt}{ChenQin} & \raisebox{5pt}{Zhang}\\
\hline
0.01 & 0.010 & 0.019 & 0.043 & 0.043 & 0.013 & 0.036 \\
0.05 & 0.049 & 0.061 & 0.083 & 0.081 & 0.036 & 0.082\\[7pt]
\end{tabular}
\begin{tabular}{@{\extracolsep{-2pt}}ccccccccccccc}
\multicolumn{12}{c}{~~~~~~~~~~~~~~~~$\boldsymbol H_1$ (shift=+0.75,--0.75)} & \!\!\!\!\!\!\!\\
 & \multicolumn{3}{c}{LRT} & \multicolumn{2}{c}{Fuji} & \multicolumn{2}{c}{Schott} & \multicolumn{2}{c}{ChenQin} & \multicolumn{2}{c}{Zhang} & \!\!\!\!\!\!\!\\[-11pt]
& \multicolumn{3}{c}{\rule{105pt}{.5pt}} & \multicolumn{2}{c}{\rule{69pt}{.5pt}} & \multicolumn{2}{c}{\rule{69pt}{.5pt}} & \multicolumn{2}{c}{\rule{69pt}{.5pt}} & \multicolumn{2}{c}{\rule{69pt}{.5pt}} & \!\!\!\!\!\!\!\\[-4pt]
$\alpha$ & exact & asymp & simul & asymp & simul & asymp & simul & asymp & simul & asymp & simul & \!\!\!\!\!\!\!\\
\hline\\[-8pt]
0.01 & 0.194 & 0.274 & 0.195 & 0.176 & 0.042 & 0.173 & 0.042 & 0.057 & 0.043 & 0.145 & 0.042 \\
0.05 & 0.427 & 0.467 & 0.429 & 0.311 & 0.202 & 0.307 & 0.202 & 0.142 & 0.190 & 0.304 & 0.219
\end{tabular}
\end{table}

\vspace{-.5cm}

\begin{table}[ht!]
\centering
\caption{\label{table7n}Normal distribution with circular covariance matrix with correlations 0.108 to 0.216 and $n_k=4,8$\\[-9pt]}
\begin{tabular}{ccccccc}
\multicolumn{7}{c}{$\boldsymbol H_0$}\\
 & \multicolumn{2}{c}{LRT} & & & \\[-1pt]
\cline{2-3}\\[-19pt]
$\alpha$ & exact & asymp & \raisebox{5pt}{Fuji} & \raisebox{5pt}{Schott} & \raisebox{5pt}{ChenQin} & \raisebox{5pt}{Zhang}\\
\hline
0.01 & 0.010 & 0.019 & 0.042 & 0.041 & 0.013 & 0.035 \\
0.05 & 0.051 & 0.062 & 0.083 & 0.082 & 0.036 & 0.081\\[7pt]
\end{tabular}
\begin{tabular}{@{\extracolsep{-2pt}}ccccccccccccc}
\multicolumn{12}{c}{~~~~~~~~~~~~~~~~$\boldsymbol H_1$ (shift=+0.5,+0.5)} & \!\!\!\!\!\!\!\\
 & \multicolumn{3}{c}{LRT} & \multicolumn{2}{c}{Fuji} & \multicolumn{2}{c}{Schott} & \multicolumn{2}{c}{ChenQin} & \multicolumn{2}{c}{Zhang} & \!\!\!\!\!\!\!\\[-11pt]
& \multicolumn{3}{c}{\rule{105pt}{.5pt}} & \multicolumn{2}{c}{\rule{69pt}{.5pt}} & \multicolumn{2}{c}{\rule{69pt}{.5pt}} & \multicolumn{2}{c}{\rule{69pt}{.5pt}} & \multicolumn{2}{c}{\rule{69pt}{.5pt}} & \!\!\!\!\!\!\!\\[-4pt]
$\alpha$ & exact & asymp & simul & asymp & simul & asymp & simul & asymp & simul & asymp & simul & \!\!\!\!\!\!\!\\
\hline\\[-8pt]
0.01 & 0.043 & 0.072 & 0.043 & 0.093 & 0.027 & 0.091 & 0.027 & 0.034 & 0.028 & 0.077 & 0.028 \\
0.05 & 0.147 & 0.171 & 0.145 & 0.166 & 0.108 & 0.163 & 0.108 & 0.079 & 0.103 & 0.163 & 0.112
\end{tabular}
\end{table}

\vspace{-.5cm}

\begin{table}[ht!]
\centering
\caption{\label{table8n}Normal distribution with circular covariance matrix with correlations 0.108 to 0.216 and $n_k=4,8$\\[-9pt]}
\begin{tabular}{ccccccc}
\multicolumn{7}{c}{$\boldsymbol H_0$}\\
 & \multicolumn{2}{c}{LRT} & & & \\[-1pt]
\cline{2-3}\\[-19pt]
$\alpha$ & exact & asymp & \raisebox{5pt}{Fuji} & \raisebox{5pt}{Schott} & \raisebox{5pt}{ChenQin} & \raisebox{5pt}{Zhang}\\
\hline
0.01 & 0.010 & 0.020 & 0.043 & 0.042 & 0.013 & 0.035 \\
0.05 & 0.051 & 0.062 & 0.084 & 0.083 & 0.035 & 0.082\\[7pt]
\end{tabular}
\begin{tabular}{@{\extracolsep{-2pt}}ccccccccccccc}
\multicolumn{12}{c}{~~~~~~~~~~~~~~~~$\boldsymbol H_1$ (shift=+0.5,--0.5)} & \!\!\!\!\!\!\!\\
 & \multicolumn{3}{c}{LRT} & \multicolumn{2}{c}{Fuji} & \multicolumn{2}{c}{Schott} & \multicolumn{2}{c}{ChenQin} & \multicolumn{2}{c}{Zhang} & \!\!\!\!\!\!\!\\[-11pt]
& \multicolumn{3}{c}{\rule{105pt}{.5pt}} & \multicolumn{2}{c}{\rule{69pt}{.5pt}} & \multicolumn{2}{c}{\rule{69pt}{.5pt}} & \multicolumn{2}{c}{\rule{69pt}{.5pt}} & \multicolumn{2}{c}{\rule{69pt}{.5pt}} & \!\!\!\!\!\!\!\\[-4pt]
$\alpha$ & exact & asymp & simul & asymp & simul & asymp & simul & asymp & simul & asymp & simul & \!\!\!\!\!\!\!\\
\hline\\[-8pt]
0.01 & 0.060 & 0.098 & 0.059 & 0.084 & 0.019 & 0.082 & 0.019 & 0.025 & 0.019 & 0.069 & 0.018 \\
0.05 & 0.188 & 0.216 & 0.186 & 0.160 & 0.098 & 0.158 & 0.098 & 0.069 & 0.096 & 0.157 & 0.101
\end{tabular}
\end{table}

\vspace{-.5cm}

\begin{table}[ht!]
\centering
\caption{\label{table9n}Normal distribution with circular covariance matrix with correlations 0.108 to 0.216 and $n_k=4,8,12$\\[-9pt]}
\begin{tabular}{ccccccc}
\multicolumn{7}{c}{$\boldsymbol H_0$}\\
 & \multicolumn{2}{c}{LRT} & & & \\[-1pt]
\cline{2-3}\\[-19pt]
$\alpha$ & exact & asymp & \raisebox{5pt}{Fuji} & \raisebox{5pt}{Schott} & \raisebox{5pt}{ChenQin} & \raisebox{5pt}{Zhang}\\
\hline
0.01 & 0.010 & 0.017 & 0.037 & 0.035 & --- & --- \\
0.05 & 0.050 & 0.058 & 0.080 & 0.077 & --- & --- \\[7pt]
\end{tabular}
\begin{tabular}{@{\extracolsep{-2pt}}ccccccccccccc}
\multicolumn{12}{c}{~~~~~~~~~~~~~~~~$\boldsymbol H_1$ (shift=+0.75,+0.75)} & \!\!\!\!\!\!\!\\
 & \multicolumn{3}{c}{LRT} & \multicolumn{2}{c}{Fuji} & \multicolumn{2}{c}{Schott} & \multicolumn{2}{c}{ChenQin} & \multicolumn{2}{c}{Zhang} & \!\!\!\!\!\!\!\\[-11pt]
& \multicolumn{3}{c}{\rule{105pt}{.5pt}} & \multicolumn{2}{c}{\rule{69pt}{.5pt}} & \multicolumn{2}{c}{\rule{69pt}{.5pt}} & \multicolumn{2}{c}{\rule{69pt}{.5pt}} & \multicolumn{2}{c}{\rule{69pt}{.5pt}} & \!\!\!\!\!\!\!\\[-4pt]
$\alpha$ & exact & asymp & simul & asymp & simul & asymp & simul & asymp & simul & asymp & simul & \!\!\!\!\!\!\!\\
\hline\\[-8pt]
0.01 & 0.317 & 0.391 & 0.313 & 0.318 & 0.143 & 0.306 & 0.143 & --- & --- & --- & --- \\
0.05 & 0.579 & 0.610 & 0.579 & 0.486 & 0.379 & 0.475 & 0.379 & --- & --- & --- & ---
\end{tabular}
\end{table}

\vspace{-.5cm}

\begin{table}[ht!]
\centering
\caption{\label{table10n}Normal distribution with circular covariance matrix with correlations 0.108 to 0.216 and $n_k=4,8,12$\\[-9pt]}
\begin{tabular}{ccccccc}
\multicolumn{7}{c}{$\boldsymbol H_0$}\\
 & \multicolumn{2}{c}{LRT} & & & \\[-1pt]
\cline{2-3}\\[-19pt]
$\alpha$ & exact & asymp & \raisebox{5pt}{Fuji} & \raisebox{5pt}{Schott} & \raisebox{5pt}{ChenQin} & \raisebox{5pt}{Zhang}\\
\hline
0.01 & 0.010 & 0.016 & 0.037 & 0.034 & --- & --- \\
0.05 & 0.050 & 0.059 & 0.080 & 0.077 & --- & --- \\[7pt]
\end{tabular}
\begin{tabular}{@{\extracolsep{-2pt}}ccccccccccccc}
\multicolumn{12}{c}{~~~~~~~~~~~~~~~~$\boldsymbol H_1$ (shift=+0.75,--0.75)} & \!\!\!\!\!\!\!\\
 & \multicolumn{3}{c}{LRT} & \multicolumn{2}{c}{Fuji} & \multicolumn{2}{c}{Schott} & \multicolumn{2}{c}{ChenQin} & \multicolumn{2}{c}{Zhang} & \!\!\!\!\!\!\!\\[-11pt]
& \multicolumn{3}{c}{\rule{105pt}{.5pt}} & \multicolumn{2}{c}{\rule{69pt}{.5pt}} & \multicolumn{2}{c}{\rule{69pt}{.5pt}} & \multicolumn{2}{c}{\rule{69pt}{.5pt}} & \multicolumn{2}{c}{\rule{69pt}{.5pt}} & \!\!\!\!\!\!\!\\[-4pt]
$\alpha$ & exact & asymp & simul & asymp & simul & asymp & simul & asymp & simul & asymp & simul & \!\!\!\!\!\!\!\\
\hline\\[-8pt]
0.01 & 0.494 & 0.574 & 0.496 & 0.312 & 0.113 & 0.299 & 0.113 & --- & --- & --- & --- \\
0.05 & 0.748 & 0.773 & 0.747 & 0.509 & 0.383 & 0.496 & 0.383 & --- & --- & --- & ---
\end{tabular}
\end{table}

\vspace{-.5cm}

\begin{table}[ht!]
\centering
\caption{\label{table11n}Normal distribution with circular covariance matrix with correlations 0.108 to 0.216 and $n_k=4,8,12$\\[-9pt]}
\begin{tabular}{ccccccc}
\multicolumn{7}{c}{$\boldsymbol H_0$}\\
 & \multicolumn{2}{c}{LRT} & & & \\[-1pt]
\cline{2-3}\\[-19pt]
$\alpha$ & exact & asymp & \raisebox{5pt}{Fuji} & \raisebox{5pt}{Schott} & \raisebox{5pt}{ChenQin} & \raisebox{5pt}{Zhang}\\
\hline
0.01 & 0.010 & 0.017 & 0.037 & 0.034 & \\
0.05 & 0.051 & 0.059 & 0.081 & 0.077 & \\[7pt]
\end{tabular}
\begin{tabular}{@{\extracolsep{-2pt}}ccccccccccccc}
\multicolumn{12}{c}{~~~~~~~~~~~~~~~~$\boldsymbol H_1$ (shift=+0.5,+0.5)} & \!\!\!\!\!\!\!\\
 & \multicolumn{3}{c}{LRT} & \multicolumn{2}{c}{Fuji} & \multicolumn{2}{c}{Schott} & \multicolumn{2}{c}{ChenQin} & \multicolumn{2}{c}{Zhang} & \!\!\!\!\!\!\!\\[-11pt]
& \multicolumn{3}{c}{\rule{105pt}{.5pt}} & \multicolumn{2}{c}{\rule{69pt}{.5pt}} & \multicolumn{2}{c}{\rule{69pt}{.5pt}} & \multicolumn{2}{c}{\rule{69pt}{.5pt}} & \multicolumn{2}{c}{\rule{69pt}{.5pt}} & \!\!\!\!\!\!\!\\[-4pt]
$\alpha$ & exact & asymp & simul & asymp & simul & asymp & simul & asymp & simul & asymp & simul & \!\!\!\!\!\!\!\\
\hline\\[-8pt]
0.01 & 0.087 & 0.123 & 0.086 & 0.124 & 0.044 & 0.118 & 0.044 & --- & --- & --- & --- \\
0.05 & 0.243 & 0.269 & 0.242 & 0.227 & 0.157 & 0.219 & 0.157 & --- & --- & --- & ---
\end{tabular}
\end{table}

\vspace{-.5cm}

\begin{table}[ht!]
\centering
\caption{\label{table12n}Normal distribution with circular covariance matrix with correlations 0.108 to 0.216 and $n_k=4,8,12$\\[-9pt]}
\begin{tabular}{ccccccc}
\multicolumn{7}{c}{$\boldsymbol H_0$}\\
 & \multicolumn{2}{c}{LRT} & & & \\[-1pt]
\cline{2-3}\\[-19pt]
$\alpha$ & exact & asymp & \raisebox{5pt}{Fuji} & \raisebox{5pt}{Schott} & \raisebox{5pt}{ChenQin} & \raisebox{5pt}{Zhang}\\
\hline
0.01 & 0.010 & 0.017 & 0.036 & 0.034 & \\
0.05 & 0.050 & 0.059 & 0.079 & 0.075 & \\[7pt]
\end{tabular}
\begin{tabular}{@{\extracolsep{-2pt}}ccccccccccccc}
\multicolumn{12}{c}{~~~~~~~~~~~~~~~~$\boldsymbol H_1$ (shift=+0.5,--0.5)} & \!\!\!\!\!\!\!\\
 & \multicolumn{3}{c}{LRT} & \multicolumn{2}{c}{Fuji} & \multicolumn{2}{c}{Schott} & \multicolumn{2}{c}{ChenQin} & \multicolumn{2}{c}{Zhang} & \!\!\!\!\!\!\!\\[-11pt]
& \multicolumn{3}{c}{\rule{105pt}{.5pt}} & \multicolumn{2}{c}{\rule{69pt}{.5pt}} & \multicolumn{2}{c}{\rule{69pt}{.5pt}} & \multicolumn{2}{c}{\rule{69pt}{.5pt}} & \multicolumn{2}{c}{\rule{69pt}{.5pt}} & \!\!\!\!\!\!\!\\[-4pt]
$\alpha$ & exact & asymp & simul & asymp & simul & asymp & simul & asymp & simul & asymp & simul & \!\!\!\!\!\!\!\\
\hline\\[-8pt]
0.01 & 0.135 & 0.182 & 0.134 & 0.109 & 0.030 & 0.102 & 0.030 & --- & --- & --- & --- \\
0.05 & 0.330 & 0.359 & 0.331 & 0.215 & 0.143 & 0.207 & 0.143 & --- & --- & --- & ---
\end{tabular}
\end{table}

\vspace{-.5cm}

\begin{table}[ht!]
\centering
\caption{\label{table13n}T distribution with 5 d.f.\ and with circular covariance matrix with correlations 0.108 to 0.216 and $n_k=4,4$\\[-9pt]}
\begin{tabular}{ccccccc}
\multicolumn{7}{c}{$\boldsymbol H_0$}\\
 & \multicolumn{2}{c}{LRT} & & & \\[-1pt]
\cline{2-3}\\[-19pt]
$\alpha$ & exact & asymp & \raisebox{5pt}{Fuji} & \raisebox{5pt}{Schott} & \raisebox{5pt}{ChenQin} & \raisebox{5pt}{Zhang}\\
\hline
0.01 & 0.007 & 0.014 & 0.022 & 0.020 & 0.054 & 0.027 \\
0.05 & 0.040 & 0.050 & 0.049 & 0.047 & 0.102 & 0.081\\[7pt]
\end{tabular}
\begin{tabular}{@{\extracolsep{-2pt}}ccccccccccccc}
\multicolumn{12}{c}{~~~~~~~~~~~~~~~~$\boldsymbol H_1$ (shift=+0.75,+0.75)} & \!\!\!\!\!\!\!\\
 & \multicolumn{3}{c}{LRT} & \multicolumn{2}{c}{Fuji} & \multicolumn{2}{c}{Schott} & \multicolumn{2}{c}{ChenQin} & \multicolumn{2}{c}{Zhang} & \!\!\!\!\!\!\!\\[-11pt]
& \multicolumn{3}{c}{\rule{105pt}{.5pt}} & \multicolumn{2}{c}{\rule{69pt}{.5pt}} & \multicolumn{2}{c}{\rule{69pt}{.5pt}} & \multicolumn{2}{c}{\rule{69pt}{.5pt}} & \multicolumn{2}{c}{\rule{69pt}{.5pt}} & \!\!\!\!\!\!\!\\[-4pt]
$\alpha$ & exact & asymp & simul & asymp & simul & asymp & simul & asymp & simul & asymp & simul & \!\!\!\!\!\!\!\\
\hline\\[-8pt]
0.01 & 0.039 & 0.067 & 0.051 & 0.063 & 0.032 & 0.060 & 0.032 & 0.106 & 0.026 & 0.076 & 0.014 \\
0.05 & 0.142 & 0.167 & 0.167 & 0.125 & 0.127 & 0.119 & 0.127 & 0.177 & 0.101 & 0.182 & 0.076
\end{tabular}
\end{table}

\vspace{-.5cm}

\begin{table}[ht!]
\centering
\caption{\label{table14n}T distribution with 5 d.f.\ and with circular covariance matrix with correlations 0.108 to 0.216 and $n_k=4,4$\\[-9pt]}
\begin{tabular}{ccccccc}
\multicolumn{7}{c}{$\boldsymbol H_0$}\\
 & \multicolumn{2}{c}{LRT} & & & \\[-1pt]
\cline{2-3}\\[-19pt]
$\alpha$ & exact & asymp & \raisebox{5pt}{Fuji} & \raisebox{5pt}{Schott} & \raisebox{5pt}{ChenQin} & \raisebox{5pt}{Zhang}\\
\hline
0.01 & 0.007 & 0.015 & 0.021 & 0.020 & 0.054 & 0.027 \\
0.05 & 0.040 & 0.050 & 0.049 & 0.047 & 0.103 & 0.081\\[7pt]
\end{tabular}
\begin{tabular}{@{\extracolsep{-2pt}}ccccccccccccc}
\multicolumn{12}{c}{~~~~~~~~~~~~~~~~$\boldsymbol H_1$ (shift=+0.75,--0.75)} & \!\!\!\!\!\!\!\\
 & \multicolumn{3}{c}{LRT} & \multicolumn{2}{c}{Fuji} & \multicolumn{2}{c}{Schott} & \multicolumn{2}{c}{ChenQin} & \multicolumn{2}{c}{Zhang} & \!\!\!\!\!\!\!\\[-11pt]
& \multicolumn{3}{c}{\rule{105pt}{.5pt}} & \multicolumn{2}{c}{\rule{69pt}{.5pt}} & \multicolumn{2}{c}{\rule{69pt}{.5pt}} & \multicolumn{2}{c}{\rule{69pt}{.5pt}} & \multicolumn{2}{c}{\rule{69pt}{.5pt}} & \!\!\!\!\!\!\!\\[-4pt]
$\alpha$ & exact & asymp & simul & asymp & simul & asymp & simul & asymp & simul & asymp & simul & \!\!\!\!\!\!\!\\
\hline\\[-8pt]
0.01 & 0.056 & 0.092 & 0.069 & 0.054 & 0.025 & 0.051 & 0.025 & 0.095 & 0.019 & 0.067 & 0.013 \\
0.05 & 0.186 & 0.215 & 0.215 & 0.117 & 0.118 & 0.112 & 0.118 & 0.171 & 0.088 & 0.179 & 0.072
\end{tabular}
\end{table}

\vspace{-.5cm}

\begin{table}[ht!]
\centering
\caption{\label{table15n}T distribution with 5 d.f.\ and with circular covariance matrix with correlations 0.108 to 0.216 and $n_k=4,8$\\[-9pt]}
\begin{tabular}{ccccccc}
\multicolumn{7}{c}{$\boldsymbol H_0$}\\
 & \multicolumn{2}{c}{LRT} & & & \\[-1pt]
\cline{2-3}\\[-19pt]
$\alpha$ & exact & asymp & \raisebox{5pt}{Fuji} & \raisebox{5pt}{Schott} & \raisebox{5pt}{ChenQin} & \raisebox{5pt}{Zhang}\\
\hline
0.01 & 0.023 & 0.038 & 0.026 & 0.025 & 0.011 & 0.020 \\
0.05 & 0.076 & 0.089 & 0.059 & 0.058 & 0.032 & 0.064\\[7pt]
\end{tabular}
\begin{tabular}{@{\extracolsep{-2pt}}ccccccccccccc}
\multicolumn{12}{c}{~~~~~~~~~~~~~~~~$\boldsymbol H_1$ (shift=+0.75,+0.75)} & \!\!\!\!\!\!\!\\
 & \multicolumn{3}{c}{LRT} & \multicolumn{2}{c}{Fuji} & \multicolumn{2}{c}{Schott} & \multicolumn{2}{c}{ChenQin} & \multicolumn{2}{c}{Zhang} & \!\!\!\!\!\!\!\\[-11pt]
& \multicolumn{3}{c}{\rule{105pt}{.5pt}} & \multicolumn{2}{c}{\rule{69pt}{.5pt}} & \multicolumn{2}{c}{\rule{69pt}{.5pt}} & \multicolumn{2}{c}{\rule{69pt}{.5pt}} & \multicolumn{2}{c}{\rule{69pt}{.5pt}} & \!\!\!\!\!\!\!\\[-4pt]
$\alpha$ & exact & asymp & simul & asymp & simul & asymp & simul & asymp & simul & asymp & simul & \!\!\!\!\!\!\!\\
\hline\\[-8pt]
0.01 & 0.097 & 0.141 & 0.048 & 0.084 & 0.039 & 0.082 & 0.039 & 0.049 & 0.045 & 0.072 & 0.016 \\
0.05 & 0.239 & 0.267 & 0.175 & 0.162 & 0.143 & 0.160 & 0.143 & 0.113 & 0.161 & 0.178 & 0.090
\end{tabular}
\end{table}

\vspace{-.5cm}

\begin{table}[ht!]
\centering
\caption{\label{table16n}T distribution with 5 d.f.\ and with circular covariance matrix with correlations 0.108 to 0.216 and $n_k=4,8$\\[-9pt]}
\begin{tabular}{ccccccc}
\multicolumn{7}{c}{$\boldsymbol H_0$}\\
 & \multicolumn{2}{c}{LRT} & & & \\[-1pt]
\cline{2-3}\\[-19pt]
$\alpha$ & exact & asymp & \raisebox{5pt}{Fuji} & \raisebox{5pt}{Schott} & \raisebox{5pt}{ChenQin} & \raisebox{5pt}{Zhang}\\
\hline
0.01 & 0.024 & 0.039 & 0.026 & 0.025 & 0.011 & 0.020 \\
0.05 & 0.077 & 0.091 & 0.061 & 0.060 & 0.032 & 0.065\\[7pt]
\end{tabular}
\begin{tabular}{@{\extracolsep{-2pt}}ccccccccccccc}
\multicolumn{12}{c}{~~~~~~~~~~~~~~~~$\boldsymbol H_1$ (shift=+0.75,--0.75)} & \!\!\!\!\!\!\!\\
 & \multicolumn{3}{c}{LRT} & \multicolumn{2}{c}{Fuji} & \multicolumn{2}{c}{Schott} & \multicolumn{2}{c}{ChenQin} & \multicolumn{2}{c}{Zhang} & \!\!\!\!\!\!\!\\[-11pt]
& \multicolumn{3}{c}{\rule{105pt}{.5pt}} & \multicolumn{2}{c}{\rule{69pt}{.5pt}} & \multicolumn{2}{c}{\rule{69pt}{.5pt}} & \multicolumn{2}{c}{\rule{69pt}{.5pt}} & \multicolumn{2}{c}{\rule{69pt}{.5pt}} & \!\!\!\!\!\!\!\\[-4pt]
$\alpha$ & exact & asymp & simul & asymp & simul & asymp & simul & asymp & simul & asymp & simul & \!\!\!\!\!\!\!\\
\hline\\[-8pt]
0.01 & 0.133 & 0.186 & 0.070 & 0.073 & 0.030 & 0.072 & 0.030 & 0.036 & 0.032 & 0.060 & 0.014 \\
0.05 & 0.299 & 0.331 & 0.225 & 0.156 & 0.132 & 0.153 & 0.132 & 0.098 & 0.144 & 0.172 & 0.080
\end{tabular}
\end{table}

\vspace{-.5cm}

\begin{table}[ht!]
\centering
\caption{\label{table17n}T distribution with 5 d.f.\ and with circular covariance matrix with correlations 0.108 to 0.216 and $n_k=4,8,12$\\[-9pt]}
\begin{tabular}{ccccccc}
\multicolumn{7}{c}{$\boldsymbol H_0$}\\
 & \multicolumn{2}{c}{LRT} & & & \\[-1pt]
\cline{2-3}\\[-19pt]
$\alpha$ & exact & asymp & \raisebox{5pt}{Fuji} & \raisebox{5pt}{Schott} & \raisebox{5pt}{ChenQin} & \raisebox{5pt}{Zhang}\\
\hline
0.01 & 0.033 & 0.045 & 0.030 & 0.028 & --- & --- \\
0.05 & 0.088 & 0.098 & 0.072 & 0.069 & --- & --- \\[7pt]
\end{tabular}
\begin{tabular}{@{\extracolsep{-2pt}}ccccccccccccc}
\multicolumn{12}{c}{~~~~~~~~~~~~~~~~$\boldsymbol H_1$ (shift=+0.75,+0.75)} & \!\!\!\!\!\!\!\\
 & \multicolumn{3}{c}{LRT} & \multicolumn{2}{c}{Fuji} & \multicolumn{2}{c}{Schott} & \multicolumn{2}{c}{ChenQin} & \multicolumn{2}{c}{Zhang} & \!\!\!\!\!\!\!\\[-11pt]
& \multicolumn{3}{c}{\rule{105pt}{.5pt}} & \multicolumn{2}{c}{\rule{69pt}{.5pt}} & \multicolumn{2}{c}{\rule{69pt}{.5pt}} & \multicolumn{2}{c}{\rule{69pt}{.5pt}} & \multicolumn{2}{c}{\rule{69pt}{.5pt}} & \!\!\!\!\!\!\!\\[-4pt]
$\alpha$ & exact & asymp & simul & asymp & simul & asymp & simul & asymp & simul & asymp & simul & \!\!\!\!\!\!\!\\
\hline\\[-8pt]
0.01 & 0.193 & 0.243 & 0.070 & 0.139 & 0.059 & 0.132 & 0.059 & --- & --- & --- & --- \\
0.05 & 0.379 & 0.404 & 0.258 & 0.257 & 0.200 & 0.249 & 0.200 & --- & --- & --- & ---
\end{tabular}
\end{table}

\vspace{-.5cm}

\begin{table}[ht!]
\centering
\caption{\label{table18n}T distribution with 5 d.f.\ and with circular covariance matrix with correlations 0.108 to 0.216 and $n_k=4,8,12$\\[-9pt]}
\begin{tabular}{ccccccc}
\multicolumn{7}{c}{$\boldsymbol H_0$}\\
 & \multicolumn{2}{c}{LRT} & & & \\[-1pt]
\cline{2-3}\\[-19pt]
$\alpha$ & exact & asymp & \raisebox{5pt}{Fuji} & \raisebox{5pt}{Schott} & \raisebox{5pt}{ChenQin} & \raisebox{5pt}{Zhang}\\
\hline
0.01 & 0.032 & 0.044 & 0.029 & 0.027 & --- & --- \\
0.05 & 0.087 & 0.097 & 0.069 & 0.066 & --- & --- \\[7pt]
\end{tabular}
\begin{tabular}{@{\extracolsep{-2pt}}ccccccccccccc}
\multicolumn{12}{c}{~~~~~~~~~~~~~~~~$\boldsymbol H_1$ (shift=+0.75,--0.75)} & \!\!\!\!\!\!\!\\
 & \multicolumn{3}{c}{LRT} & \multicolumn{2}{c}{Fuji} & \multicolumn{2}{c}{Schott} & \multicolumn{2}{c}{ChenQin} & \multicolumn{2}{c}{Zhang} & \!\!\!\!\!\!\!\\[-11pt]
& \multicolumn{3}{c}{\rule{105pt}{.5pt}} & \multicolumn{2}{c}{\rule{69pt}{.5pt}} & \multicolumn{2}{c}{\rule{69pt}{.5pt}} & \multicolumn{2}{c}{\rule{69pt}{.5pt}} & \multicolumn{2}{c}{\rule{69pt}{.5pt}} & \!\!\!\!\!\!\!\\[-4pt]
$\alpha$ & exact & asymp & simul & asymp & simul & asymp & simul & asymp & simul & asymp & simul & \!\!\!\!\!\!\!\\
\hline\\[-8pt]
0.01 & 0.278 & 0.338 & 0.116 & 0.124 & 0.048 & 0.116 & 0.048 & --- & --- & --- & --- \\
0.05 & 0.489 & 0.515 & 0.365 & 0.248 & 0.193 & 0.239 & 0.193 & --- & --- & --- & ---
\end{tabular}
\end{table}

\vspace{-.5cm}

\begin{table}[ht!]
\centering
\caption{\label{table19n}T distribution with 3 d.f.\ and with circular covariance matrix with correlations 0.108 to 0.216 and $n_k=4,4$\\[-9pt]}
\begin{tabular}{ccccccc}
\multicolumn{7}{c}{$\boldsymbol H_0$}\\
 & \multicolumn{2}{c}{LRT} & & & \\[-1pt]
\cline{2-3}\\[-19pt]
$\alpha$ & exact & asymp & \raisebox{5pt}{Fuji} & \raisebox{5pt}{Schott} & \raisebox{5pt}{ChenQin} & \raisebox{5pt}{Zhang}\\
\hline
0.01 & 0.005 & 0.012 & 0.013 & 0.012 & 0.053 & 0.020 \\
0.05 & 0.034 & 0.043 & 0.032 & 0.031 & 0.103 & 0.089 \\[7pt]
\end{tabular}
\begin{tabular}{@{\extracolsep{-2pt}}ccccccccccccc}
\multicolumn{12}{c}{~~~~~~~~~~~~~~~~$\boldsymbol H_1$ (shift=+0.75,+0.75)} & \!\!\!\!\!\!\!\\
 & \multicolumn{3}{c}{LRT} & \multicolumn{2}{c}{Fuji} & \multicolumn{2}{c}{Schott} & \multicolumn{2}{c}{ChenQin} & \multicolumn{2}{c}{Zhang} & \!\!\!\!\!\!\!\\[-11pt]
& \multicolumn{3}{c}{\rule{105pt}{.5pt}} & \multicolumn{2}{c}{\rule{69pt}{.5pt}} & \multicolumn{2}{c}{\rule{69pt}{.5pt}} & \multicolumn{2}{c}{\rule{69pt}{.5pt}} & \multicolumn{2}{c}{\rule{69pt}{.5pt}} & \!\!\!\!\!\!\!\\[-4pt]
$\alpha$ & exact & asymp & simul & asymp & simul & asymp & simul & asymp & simul & asymp & simul & \!\!\!\!\!\!\!\\
\hline\\[-8pt]
0.01 & 0.028 & 0.050 & 0.045 & 0.036 & 0.028 & 0.034 & 0.028 & 0.095 & 0.024 & 0.050 & 0.010 \\
0.05 & 0.110 & 0.131 & 0.145 & 0.078 & 0.113 & 0.075 & 0.113 & 0.165 & 0.090 & 0.166 & 0.056
\end{tabular}
\end{table}

\vspace{-.5cm}

\begin{table}[ht!]
\centering
\caption{\label{table20n}T distribution with 3 d.f.\ and with circular covariance matrix with correlations 0.108 to 0.216 and $n_k=4,4$\\[-9pt]}
\begin{tabular}{ccccccc}
\multicolumn{7}{c}{$\boldsymbol H_0$}\\
 & \multicolumn{2}{c}{LRT} & & & \\[-1pt]
\cline{2-3}\\[-19pt]
$\alpha$ & exact & asymp & \raisebox{5pt}{Fuji} & \raisebox{5pt}{Schott} & \raisebox{5pt}{ChenQin} & \raisebox{5pt}{Zhang}\\
\hline
0.01 & 0.005 & 0.011 & 0.013 & 0.012 & 0.054 & 0.019 \\
0.05 & 0.033 & 0.043 & 0.032 & 0.030 & 0.105 & 0.091 \\[7pt]
\end{tabular}
\begin{tabular}{@{\extracolsep{-2pt}}ccccccccccccc}
\multicolumn{12}{c}{~~~~~~~~~~~~~~~~$\boldsymbol H_1$ (shift=+0.75,--0.75)} & \!\!\!\!\!\!\!\\
 & \multicolumn{3}{c}{LRT} & \multicolumn{2}{c}{Fuji} & \multicolumn{2}{c}{Schott} & \multicolumn{2}{c}{ChenQin} & \multicolumn{2}{c}{Zhang} & \!\!\!\!\!\!\!\\[-11pt]
& \multicolumn{3}{c}{\rule{105pt}{.5pt}} & \multicolumn{2}{c}{\rule{69pt}{.5pt}} & \multicolumn{2}{c}{\rule{69pt}{.5pt}} & \multicolumn{2}{c}{\rule{69pt}{.5pt}} & \multicolumn{2}{c}{\rule{69pt}{.5pt}} & \!\!\!\!\!\!\!\\[-4pt]
$\alpha$ & exact & asymp & simul & asymp & simul & asymp & simul & asymp & simul & asymp & simul & \!\!\!\!\!\!\!\\
\hline\\[-8pt]
0.01 & 0.038 & 0.066 & 0.061 & 0.030 & 0.024 & 0.028 & 0.024 & 0.085 & 0.017 & 0.042 & 0.011 \\
0.05 & 0.141 & 0.164 & 0.182 & 0.071 & 0.106 & 0.067 & 0.106 & 0.158 & 0.080 & 0.161 & 0.055
\end{tabular}
\end{table}

\vspace{-.5cm}

\begin{table}[ht!]
\centering
\caption{\label{table21n}T distribution with 3 d.f.\ and with circular covariance matrix with correlations 0.108 to 0.216 and $n_k=4,8$\\[-9pt]}
\begin{tabular}{ccccccc}
\multicolumn{7}{c}{$\boldsymbol H_0$}\\
 & \multicolumn{2}{c}{LRT} & & & \\[-1pt]
\cline{2-3}\\[-19pt]
$\alpha$ & exact & asymp & \raisebox{5pt}{Fuji} & \raisebox{5pt}{Schott} & \raisebox{5pt}{ChenQin} & \raisebox{5pt}{Zhang}\\
\hline
0.01 & 0.040 & 0.059 & 0.019 & 0.018 & 0.010 & 0.016 \\
0.05 & 0.104 & 0.117 & 0.048 & 0.047 & 0.030 & 0.067 \\[7pt]
\end{tabular}
\begin{tabular}{@{\extracolsep{-2pt}}ccccccccccccc}
\multicolumn{12}{c}{~~~~~~~~~~~~~~~~$\boldsymbol H_1$ (shift=+0.75,+0.75)} & \!\!\!\!\!\!\!\\
 & \multicolumn{3}{c}{LRT} & \multicolumn{2}{c}{Fuji} & \multicolumn{2}{c}{Schott} & \multicolumn{2}{c}{ChenQin} & \multicolumn{2}{c}{Zhang} & \!\!\!\!\!\!\!\\[-11pt]
& \multicolumn{3}{c}{\rule{105pt}{.5pt}} & \multicolumn{2}{c}{\rule{69pt}{.5pt}} & \multicolumn{2}{c}{\rule{69pt}{.5pt}} & \multicolumn{2}{c}{\rule{69pt}{.5pt}} & \multicolumn{2}{c}{\rule{69pt}{.5pt}} & \!\!\!\!\!\!\!\\[-4pt]
$\alpha$ & exact & asymp & simul & asymp & simul & asymp & simul & asymp & simul & asymp & simul & \!\!\!\!\!\!\!\\
\hline\\[-8pt]
0.01 & 0.096 & 0.134 & 0.029 & 0.050 & 0.030 & 0.048 & 0.030 & 0.036 & 0.037 & 0.043 & 0.011 \\
0.05 & 0.216 & 0.239 & 0.116 & 0.107 & 0.111 & 0.105 & 0.111 & 0.090 & 0.134 & 0.144 & 0.058
\end{tabular}
\end{table}

\vspace{-.5cm}

\begin{table}[ht!]
\centering
\caption{\label{table22n}T distribution with 3 d.f.\ and with circular covariance matrix with correlations 0.108 to 0.216 and $n_k=4,8$\\[-9pt]}
\begin{tabular}{ccccccc}
\multicolumn{7}{c}{$\boldsymbol H_0$}\\
 & \multicolumn{2}{c}{LRT} & & & \\[-1pt]
\cline{2-3}\\[-19pt]
$\alpha$ & exact & asymp & \raisebox{5pt}{Fuji} & \raisebox{5pt}{Schott} & \raisebox{5pt}{ChenQin} & \raisebox{5pt}{Zhang}\\
\hline
0.01 & 0.041 & 0.060 & 0.019 & 0.018 & 0.009 & 0.015 \\
0.05 & 0.104 & 0.117 & 0.048 & 0.047 & 0.030 & 0.068 \\[7pt]
\end{tabular}
\begin{tabular}{@{\extracolsep{-2pt}}ccccccccccccc}
\multicolumn{12}{c}{~~~~~~~~~~~~~~~~$\boldsymbol H_1$ (shift=+0.75,--0.75)} & \!\!\!\!\!\!\!\\
 & \multicolumn{3}{c}{LRT} & \multicolumn{2}{c}{Fuji} & \multicolumn{2}{c}{Schott} & \multicolumn{2}{c}{ChenQin} & \multicolumn{2}{c}{Zhang} & \!\!\!\!\!\!\!\\[-11pt]
& \multicolumn{3}{c}{\rule{105pt}{.5pt}} & \multicolumn{2}{c}{\rule{69pt}{.5pt}} & \multicolumn{2}{c}{\rule{69pt}{.5pt}} & \multicolumn{2}{c}{\rule{69pt}{.5pt}} & \multicolumn{2}{c}{\rule{69pt}{.5pt}} & \!\!\!\!\!\!\!\\[-4pt]
$\alpha$ & exact & asymp & simul & asymp & simul & asymp & simul & asymp & simul & asymp & simul & \!\!\!\!\!\!\!\\
\hline\\[-8pt]
0.01 & 0.119 & 0.164 & 0.037 & 0.043 & 0.024 & 0.042 & 0.024 & 0.026 & 0.028 & 0.036 & 0.010 \\
0.05 & 0.253 & 0.278 & 0.141 & 0.101 & 0.104 & 0.100 & 0.104 & 0.076 & 0.121 & 0.136 & 0.057
\end{tabular}
\end{table}

\vspace{-.5cm}

\begin{table}[ht!]
\centering
\caption{\label{table23n}T distribution with 3 d.f.\ and with circular covariance matrix with correlations 0.108 to 0.216 and $n_k=4,8,12$\\[-9pt]}
\begin{tabular}{ccccccc}
\multicolumn{7}{c}{$\boldsymbol H_0$}\\
 & \multicolumn{2}{c}{LRT} & & & \\[-1pt]
\cline{2-3}\\[-19pt]
$\alpha$ & exact & asymp & \raisebox{5pt}{Fuji} & \raisebox{5pt}{Schott} & \raisebox{5pt}{ChenQin} & \raisebox{5pt}{Zhang}\\
\hline
0.01 & 0.061 & 0.074 & 0.027 & 0.025 & --- & --- \\
0.05 & 0.118 & 0.128 & 0.070 & 0.067 & --- & --- \\[7pt]
\end{tabular}
\begin{tabular}{@{\extracolsep{-2pt}}ccccccccccccc}
\multicolumn{12}{c}{~~~~~~~~~~~~~~~~$\boldsymbol H_1$ (shift=+0.75,+0.75)} & \!\!\!\!\!\!\!\\
 & \multicolumn{3}{c}{LRT} & \multicolumn{2}{c}{Fuji} & \multicolumn{2}{c}{Schott} & \multicolumn{2}{c}{ChenQin} & \multicolumn{2}{c}{Zhang} & \!\!\!\!\!\!\!\\[-11pt]
& \multicolumn{3}{c}{\rule{105pt}{.5pt}} & \multicolumn{2}{c}{\rule{69pt}{.5pt}} & \multicolumn{2}{c}{\rule{69pt}{.5pt}} & \multicolumn{2}{c}{\rule{69pt}{.5pt}} & \multicolumn{2}{c}{\rule{69pt}{.5pt}} & \!\!\!\!\!\!\!\\[-4pt]
$\alpha$ & exact & asymp & simul & asymp & simul & asymp & simul & asymp & simul & asymp & simul & \!\!\!\!\!\!\!\\
\hline\\[-8pt]
0.01 & 0.159 & 0.194 & 0.022 & 0.080 & 0.036 & 0.075 & 0.036 & --- & --- & --- & --- \\
0.05 & 0.294 & 0.314 & 0.132 & 0.168 & 0.129 & 0.161 & 0.129 & --- & --- & --- & ---
\end{tabular}
\end{table}

\vspace{-.5cm}

\begin{table}[ht!]
\centering
\caption{\label{table24n}T distribution with 3 d.f.\ and with circular covariance matrix with correlations 0.108 to 0.216 and $n_k=4,8,12$\\[-9pt]}
\begin{tabular}{ccccccc}
\multicolumn{7}{c}{$\boldsymbol H_0$}\\
 & \multicolumn{2}{c}{LRT} & & & \\[-1pt]
\cline{2-3}\\[-19pt]
$\alpha$ & exact & asymp & \raisebox{5pt}{Fuji} & \raisebox{5pt}{Schott} & \raisebox{5pt}{ChenQin} & \raisebox{5pt}{Zhang}\\
\hline
0.01 & 0.060 & 0.074 & 0.027 & 0.025 & --- & --- \\
0.05 & 0.117 & 0.126 & 0.069 & 0.066 & --- & --- \\[7pt]
\end{tabular}
\begin{tabular}{@{\extracolsep{-2pt}}ccccccccccccc}
\multicolumn{12}{c}{~~~~~~~~~~~~~~~~$\boldsymbol H_1$ (shift=+0.75,--0.75)} & \!\!\!\!\!\!\!\\
 & \multicolumn{3}{c}{LRT} & \multicolumn{2}{c}{Fuji} & \multicolumn{2}{c}{Schott} & \multicolumn{2}{c}{ChenQin} & \multicolumn{2}{c}{Zhang} & \!\!\!\!\!\!\!\\[-11pt]
& \multicolumn{3}{c}{\rule{105pt}{.5pt}} & \multicolumn{2}{c}{\rule{69pt}{.5pt}} & \multicolumn{2}{c}{\rule{69pt}{.5pt}} & \multicolumn{2}{c}{\rule{69pt}{.5pt}} & \multicolumn{2}{c}{\rule{69pt}{.5pt}} & \!\!\!\!\!\!\!\\[-4pt]
$\alpha$ & exact & asymp & simul & asymp & simul & asymp & simul & asymp & simul & asymp & simul & \!\!\!\!\!\!\!\\
\hline\\[-8pt]
0.01 & 0.203 & 0.244 & 0.030 & 0.071 & 0.028 & 0.066 & 0.028 & --- & --- & --- & --- \\
0.05 & 0.357 & 0.377 & 0.168 & 0.158 & 0.118 & 0.152 & 0.118 & --- & --- & --- & ---
\end{tabular}
\end{table}

\vspace{-.5cm}

\begin{table}[ht!]
\centering
\caption{\label{table25n}Cauchy distribution with circular covariance matrix with correlations 0.108 to 0.216 and $n_k=4,4$\\[-9pt]}
\begin{tabular}{ccccccc}
\multicolumn{7}{c}{$\boldsymbol H_0$}\\
 & \multicolumn{2}{c}{LRT} & & & \\[-1pt]
\cline{2-3}\\[-19pt]
$\alpha$ & exact & asymp & \raisebox{5pt}{Fuji} & \raisebox{5pt}{Schott} & \raisebox{5pt}{ChenQin} & \raisebox{5pt}{Zhang}\\
\hline
0.01 & 0.002 & 0.004 & 0.002 & 0.001 & 0.042 & 0.157 \\
0.05 & 0.013 & 0.017 & 0.006 & 0.005 & 0.099 & 0.371 \\[7pt]
\end{tabular}
\begin{tabular}{@{\extracolsep{-2pt}}ccccccccccccc}
\multicolumn{12}{c}{~~~~~~~~~~~~~~~~$\boldsymbol H_1$ (shift=+1.5,+1.5)} & \!\!\!\!\!\!\!\\
 & \multicolumn{3}{c}{LRT} & \multicolumn{2}{c}{Fuji} & \multicolumn{2}{c}{Schott} & \multicolumn{2}{c}{ChenQin} & \multicolumn{2}{c}{Zhang} & \!\!\!\!\!\!\!\\[-11pt]
& \multicolumn{3}{c}{\rule{105pt}{.5pt}} & \multicolumn{2}{c}{\rule{69pt}{.5pt}} & \multicolumn{2}{c}{\rule{69pt}{.5pt}} & \multicolumn{2}{c}{\rule{69pt}{.5pt}} & \multicolumn{2}{c}{\rule{69pt}{.5pt}} & \!\!\!\!\!\!\!\\[-4pt]
$\alpha$ & exact & asymp & simul & asymp & simul & asymp & simul & asymp & simul & asymp & simul & \!\!\!\!\!\!\!\\
\hline\\[-8pt]
0.01 & 0.034 & 0.052 & 0.091 & 0.021 & 0.057 & 0.019 & 0.057 & 0.107 & 0.044 & 0.194 & 0.010 \\
0.05 & 0.101 & 0.117 & 0.203 & 0.042 & 0.149 & 0.040 & 0.149 & 0.186 & 0.120 & 0.460 & 0.050
\end{tabular}
\end{table}

\vspace{-.5cm}

\begin{table}[ht!]
\centering
\caption{\label{table26n}Cauchy distribution with circular covariance matrix with correlations 0.108 to 0.216 and $n_k=4,4$\\[-9pt]}
\begin{tabular}{ccccccc}
\multicolumn{7}{c}{$\boldsymbol H_0$}\\
 & \multicolumn{2}{c}{LRT} & & & \\[-1pt]
\cline{2-3}\\[-19pt]
$\alpha$ & exact & asymp & \raisebox{5pt}{Fuji} & \raisebox{5pt}{Schott} & \raisebox{5pt}{ChenQin} & \raisebox{5pt}{Zhang}\\
\hline
0.01 & 0.002 & 0.004 & 0.002 & 0.002 & 0.042 & 0.158 \\
0.05 & 0.013 & 0.017 & 0.006 & 0.006 & 0.100 & 0.370 \\[7pt]
\end{tabular}
\begin{tabular}{@{\extracolsep{-2pt}}ccccccccccccc}
\multicolumn{12}{c}{~~~~~~~~~~~~~~~~$\boldsymbol H_1$ (shift=+1.5,--1.5)} & \!\!\!\!\!\!\!\\
 & \multicolumn{3}{c}{LRT} & \multicolumn{2}{c}{Fuji} & \multicolumn{2}{c}{Schott} & \multicolumn{2}{c}{ChenQin} & \multicolumn{2}{c}{Zhang} & \!\!\!\!\!\!\!\\[-11pt]
& \multicolumn{3}{c}{\rule{105pt}{.5pt}} & \multicolumn{2}{c}{\rule{69pt}{.5pt}} & \multicolumn{2}{c}{\rule{69pt}{.5pt}} & \multicolumn{2}{c}{\rule{69pt}{.5pt}} & \multicolumn{2}{c}{\rule{69pt}{.5pt}} & \!\!\!\!\!\!\!\\[-4pt]
$\alpha$ & exact & asymp & simul & asymp & simul & asymp & simul & asymp & simul & asymp & simul & \!\!\!\!\!\!\!\\
\hline\\[-8pt]
0.01 & 0.051 & 0.076 & 0.117 & 0.018 & 0.054 & 0.017 & 0.054 & 0.096 & 0.034 & 0.189 & 0.010 \\
0.05 & 0.133 & 0.149 & 0.237 & 0.039 & 0.149 & 0.038 & 0.149 & 0.178 & 0.110 & 0.463 & 0.050
\end{tabular}
\end{table}

\vspace{-.5cm}

\begin{table}[ht!]
\centering
\caption{\label{table27n}Cauchy distribution with circular covariance matrix with correlations 0.108 to 0.216 and $n_k=4,8$\\[-9pt]}
\begin{tabular}{ccccccc}
\multicolumn{7}{c}{$\boldsymbol H_0$}\\
 & \multicolumn{2}{c}{LRT} & & & \\[-1pt]
\cline{2-3}\\[-19pt]
$\alpha$ & exact & asymp & \raisebox{5pt}{Fuji} & \raisebox{5pt}{Schott} & \raisebox{5pt}{ChenQin} & \raisebox{5pt}{Zhang}\\
\hline
0.01 & 0.139 & 0.165 & 0.006 & 0.006 & 0.003 & 0.163 \\
0.05 & 0.209 & 0.219 & 0.020 & 0.019 & 0.017 & 0.345 \\[7pt]
\end{tabular}
\begin{tabular}{@{\extracolsep{-2pt}}ccccccccccccc}
\multicolumn{12}{c}{~~~~~~~~~~~~~~~~$\boldsymbol H_1$ (shift=+1.5,+1.5)} & \!\!\!\!\!\!\!\\
 & \multicolumn{3}{c}{LRT} & \multicolumn{2}{c}{Fuji} & \multicolumn{2}{c}{Schott} & \multicolumn{2}{c}{ChenQin} & \multicolumn{2}{c}{Zhang} & \!\!\!\!\!\!\!\\[-11pt]
& \multicolumn{3}{c}{\rule{105pt}{.5pt}} & \multicolumn{2}{c}{\rule{69pt}{.5pt}} & \multicolumn{2}{c}{\rule{69pt}{.5pt}} & \multicolumn{2}{c}{\rule{69pt}{.5pt}} & \multicolumn{2}{c}{\rule{69pt}{.5pt}} & \!\!\!\!\!\!\!\\[-4pt]
$\alpha$ & exact & asymp & simul & asymp & simul & asymp & simul & asymp & simul & asymp & simul & \!\!\!\!\!\!\!\\
\hline\\[-8pt]
0.01 & 0.197 & 0.230 & 0.023 & 0.022 & 0.033 & 0.022 & 0.033 & 0.040 & 0.069 & 0.188 & 0.010 \\
0.05 & 0.288 & 0.302 & 0.083 & 0.053 & 0.102 & 0.052 & 0.102 & 0.090 & 0.160 & 0.414 & 0.050
\end{tabular}
\end{table}

\vspace{-.5cm}

\begin{table}[ht!]
\centering
\caption{\label{table28n}Cauchy distribution with circular covariance matrix with correlations 0.108 to 0.216 and $n_k=4,8$\\[-9pt]}
\begin{tabular}{ccccccc}
\multicolumn{7}{c}{$\boldsymbol H_0$}\\
 & \multicolumn{2}{c}{LRT} & & & \\[-1pt]
\cline{2-3}\\[-19pt]
$\alpha$ & exact & asymp & \raisebox{5pt}{Fuji} & \raisebox{5pt}{Schott} & \raisebox{5pt}{ChenQin} & \raisebox{5pt}{Zhang}\\
\hline
0.01 & 0.141 & 0.168 & 0.006 & 0.006 & 0.003 & 0.163 \\
0.05 & 0.212 & 0.223 & 0.020 & 0.020 & 0.016 & 0.344 \\[7pt]
\end{tabular}
\begin{tabular}{@{\extracolsep{-2pt}}ccccccccccccc}
\multicolumn{12}{c}{~~~~~~~~~~~~~~~~$\boldsymbol H_1$ (shift=+1.5,--1.5)} & \!\!\!\!\!\!\!\\
 & \multicolumn{3}{c}{LRT} & \multicolumn{2}{c}{Fuji} & \multicolumn{2}{c}{Schott} & \multicolumn{2}{c}{ChenQin} & \multicolumn{2}{c}{Zhang} & \!\!\!\!\!\!\!\\[-11pt]
& \multicolumn{3}{c}{\rule{105pt}{.5pt}} & \multicolumn{2}{c}{\rule{69pt}{.5pt}} & \multicolumn{2}{c}{\rule{69pt}{.5pt}} & \multicolumn{2}{c}{\rule{69pt}{.5pt}} & \multicolumn{2}{c}{\rule{69pt}{.5pt}} & \!\!\!\!\!\!\!\\[-4pt]
$\alpha$ & exact & asymp & simul & asymp & simul & asymp & simul & asymp & simul & asymp & simul & \!\!\!\!\!\!\!\\
\hline\\[-8pt]
0.01 & 0.217 & 0.255 & 0.030 & 0.021 & 0.031 & 0.020 & 0.031 & 0.030 & 0.063 & 0.185 & 0.010 \\
0.05 & 0.313 & 0.328 & 0.092 & 0.053 & 0.101 & 0.051 & 0.101 & 0.080 & 0.155 & 0.413 & 0.050
\end{tabular}
\end{table}

\vspace{-.5cm}

\begin{table}[ht!]
\centering
\caption{\label{table29n}Cauchy distribution with circular covariance matrix with correlations 0.108 to 0.216 and $n_k=4,8,12$\\[-9pt]}
\begin{tabular}{ccccccc}
\multicolumn{7}{c}{$\boldsymbol H_0$}\\
 & \multicolumn{2}{c}{LRT} & & & \\[-1pt]
\cline{2-3}\\[-19pt]
$\alpha$ & exact & asymp & \raisebox{5pt}{Fuji} & \raisebox{5pt}{Schott} & \raisebox{5pt}{ChenQin} & \raisebox{5pt}{Zhang}\\
\hline
0.01 & 0.148 & 0.158 & 0.015 & 0.014 & --- & --- \\
0.05 & 0.185 & 0.190 & 0.101 & 0.097 & --- & --- \\[7pt]
\end{tabular}
\begin{tabular}{@{\extracolsep{-2pt}}ccccccccccccc}
\multicolumn{12}{c}{~~~~~~~~~~~~~~~~$\boldsymbol H_1$ (shift=+2.0,+2.0)} & \!\!\!\!\!\!\!\\
 & \multicolumn{3}{c}{LRT} & \multicolumn{2}{c}{Fuji} & \multicolumn{2}{c}{Schott} & \multicolumn{2}{c}{ChenQin} & \multicolumn{2}{c}{Zhang} & \!\!\!\!\!\!\!\\[-11pt]
& \multicolumn{3}{c}{\rule{105pt}{.5pt}} & \multicolumn{2}{c}{\rule{69pt}{.5pt}} & \multicolumn{2}{c}{\rule{69pt}{.5pt}} & \multicolumn{2}{c}{\rule{69pt}{.5pt}} & \multicolumn{2}{c}{\rule{69pt}{.5pt}} & \!\!\!\!\!\!\!\\[-4pt]
$\alpha$ & exact & asymp & simul & asymp & simul & asymp & simul & asymp & simul & asymp & simul & \!\!\!\!\!\!\!\\
\hline\\[-8pt]
0.01 & 0.227 & 0.245 & 0.015 & 0.046 & 0.034 & 0.042 & 0.034 & --- & --- & --- & --- \\
0.05 & 0.291 & 0.299 & 0.065 & 0.150 & 0.095 & 0.145 & 0.095 & --- & --- & --- & ---
\end{tabular}
\end{table}

\vspace{-.5cm}

\begin{table}[ht!]
\centering
\caption{\label{table30n}Cauchy distribution with circular covariance matrix with correlations 0.108 to 0.216 and $n_k=4,8,12$\\[-9pt]}
\begin{tabular}{ccccccc}
\multicolumn{7}{c}{$\boldsymbol H_0$}\\
 & \multicolumn{2}{c}{LRT} & & & \\[-1pt]
\cline{2-3}\\[-19pt]
$\alpha$ & exact & asymp & \raisebox{5pt}{Fuji} & \raisebox{5pt}{Schott} & \raisebox{5pt}{ChenQin} & \raisebox{5pt}{Zhang}\\
\hline
0.01 & 0.148 & 0.158 & 0.015 & 0.014 & --- & --- \\
0.05 & 0.185 & 0.191 & 0.101 & 0.097 & --- & --- \\[7pt]
\end{tabular}
\begin{tabular}{@{\extracolsep{-2pt}}ccccccccccccc}
\multicolumn{12}{c}{~~~~~~~~~~~~~~~~$\boldsymbol H_1$ (shift=+2.0,--2.0)} & \!\!\!\!\!\!\!\\
 & \multicolumn{3}{c}{LRT} & \multicolumn{2}{c}{Fuji} & \multicolumn{2}{c}{Schott} & \multicolumn{2}{c}{ChenQin} & \multicolumn{2}{c}{Zhang} & \!\!\!\!\!\!\!\\[-11pt]
& \multicolumn{3}{c}{\rule{105pt}{.5pt}} & \multicolumn{2}{c}{\rule{69pt}{.5pt}} & \multicolumn{2}{c}{\rule{69pt}{.5pt}} & \multicolumn{2}{c}{\rule{69pt}{.5pt}} & \multicolumn{2}{c}{\rule{69pt}{.5pt}} & \!\!\!\!\!\!\!\\[-4pt]
$\alpha$ & exact & asymp & simul & asymp & simul & asymp & simul & asymp & simul & asymp & simul & \!\!\!\!\!\!\!\\
\hline\\[-8pt]
0.01 & 0.248 & 0.268 & 0.015 & 0.043 & 0.031 & 0.040 & 0.031 & --- & --- & --- & --- \\
0.05 & 0.317 & 0.325 & 0.070 & 0.149 & 0.092 & 0.144 & 0.092 & --- & --- & --- & ---
\end{tabular}
\end{table}

\vspace{-.5cm}

\begin{table}[ht!]
\centering
\caption{\label{table31n}Skew Normal distribution with circular covariance matrix with correlations 0.108 to 0.216 and $n_k=4,4$\\[-9pt]}
\begin{tabular}{ccccccc}
\multicolumn{7}{c}{$\boldsymbol H_0$}\\
 & \multicolumn{2}{c}{LRT} & & & \\[-1pt]
\cline{2-3}\\[-19pt]
$\alpha$ & exact & asymp & \raisebox{5pt}{Fuji} & \raisebox{5pt}{Schott} & \raisebox{5pt}{ChenQin} & \raisebox{5pt}{Zhang}\\
\hline
0.01 & 0.010 & 0.020 & 0.037 & 0.034 & 0.005 & 0.041 \\
0.05 & 0.051 & 0.063 & 0.086 & 0.082 & 0.013 & 0.100 \\[7pt]
\end{tabular}
\begin{tabular}{@{\extracolsep{-2pt}}ccccccccccccc}
\multicolumn{12}{c}{~~~~~~~~~~~~~~~~$\boldsymbol H_1$ (shift=+0.75,+0.75)} & \!\!\!\!\!\!\!\\
 & \multicolumn{3}{c}{LRT} & \multicolumn{2}{c}{Fuji} & \multicolumn{2}{c}{Schott} & \multicolumn{2}{c}{ChenQin} & \multicolumn{2}{c}{Zhang} & \!\!\!\!\!\!\!\\[-11pt]
& \multicolumn{3}{c}{\rule{105pt}{.5pt}} & \multicolumn{2}{c}{\rule{69pt}{.5pt}} & \multicolumn{2}{c}{\rule{69pt}{.5pt}} & \multicolumn{2}{c}{\rule{69pt}{.5pt}} & \multicolumn{2}{c}{\rule{69pt}{.5pt}} & \!\!\!\!\!\!\!\\[-4pt]
$\alpha$ & exact & asymp & simul & asymp & simul & asymp & simul & asymp & simul & asymp & simul & \!\!\!\!\!\!\!\\
\hline\\[-8pt]
0.01 & 0.074 & 0.118 & 0.073 & 0.201 & 0.086 & 0.191 & 0.086 & 0.011 & 0.020 & 0.202 & 0.109 \\
0.05 & 0.226 & 0.258 & 0.223 & 0.338 & 0.245 & 0.329 & 0.245 & 0.025 & 0.070 & 0.358 & 0.281
\end{tabular}
\end{table}

\vspace{-.5cm}

\begin{table}[ht!]
\centering
\caption{\label{table32n}Skew Normal distribution with circular covariance matrix with correlations 0.108 to 0.216 and $n_k=4,4$\\[-9pt]}
\begin{tabular}{ccccccc}
\multicolumn{7}{c}{$\boldsymbol H_0$}\\
 & \multicolumn{2}{c}{LRT} & & & \\[-1pt]
\cline{2-3}\\[-19pt]
$\alpha$ & exact & asymp & \raisebox{5pt}{Fuji} & \raisebox{5pt}{Schott} & \raisebox{5pt}{ChenQin} & \raisebox{5pt}{Zhang}\\
\hline
0.01 & 0.010 & 0.019 & 0.036 & 0.033 & 0.005 & 0.040 \\
0.05 & 0.049 & 0.061 & 0.083 & 0.078 & 0.013 & 0.096 \\[7pt]
\end{tabular}
\begin{tabular}{@{\extracolsep{-2pt}}ccccccccccccc}
\multicolumn{12}{c}{~~~~~~~~~~~~~~~~$\boldsymbol H_1$ (shift=+0.75,--0.75)} & \!\!\!\!\!\!\!\\
 & \multicolumn{3}{c}{LRT} & \multicolumn{2}{c}{Fuji} & \multicolumn{2}{c}{Schott} & \multicolumn{2}{c}{ChenQin} & \multicolumn{2}{c}{Zhang} & \!\!\!\!\!\!\!\\[-11pt]
& \multicolumn{3}{c}{\rule{105pt}{.5pt}} & \multicolumn{2}{c}{\rule{69pt}{.5pt}} & \multicolumn{2}{c}{\rule{69pt}{.5pt}} & \multicolumn{2}{c}{\rule{69pt}{.5pt}} & \multicolumn{2}{c}{\rule{69pt}{.5pt}} & \!\!\!\!\!\!\!\\[-4pt]
$\alpha$ & exact & asymp & simul & asymp & simul & asymp & simul & asymp & simul & asymp & simul & \!\!\!\!\!\!\!\\
\hline\\[-8pt]
0.01 & 0.097 & 0.151 & 0.097 & 0.193 & 0.076 & 0.183 & 0.076 & 0.028 & 0.045 & 0.193 & 0.093 \\
0.05 & 0.272 & 0.308 & 0.274 & 0.333 & 0.241 & 0.323 & 0.241 & 0.056 & 0.148 & 0.355 & 0.281
\end{tabular}
\end{table}

\vspace{-.5cm}

\begin{table}[ht!]
\centering
\caption{\label{table33n}Skew Normal distribution with circular covariance matrix with correlations 0.108 to 0.216 and $n_k=4,8$\\[-9pt]}
\begin{tabular}{ccccccc}
\multicolumn{7}{c}{$\boldsymbol H_0$}\\
 & \multicolumn{2}{c}{LRT} & & & \\[-1pt]
\cline{2-3}\\[-19pt]
$\alpha$ & exact & asymp & \raisebox{5pt}{Fuji} & \raisebox{5pt}{Schott} & \raisebox{5pt}{ChenQin} & \raisebox{5pt}{Zhang}\\
\hline
0.01 & 0.010 & 0.019 & 0.028 & 0.028 & 0.003 & 0.023 \\
0.05 & 0.050 & 0.061 & 0.073 & 0.071 & 0.012 & 0.072 \\[7pt]
\end{tabular}
\begin{tabular}{@{\extracolsep{-2pt}}ccccccccccccc}
\multicolumn{12}{c}{~~~~~~~~~~~~~~~~$\boldsymbol H_1$ (shift=+0.75,+0.75)} & \!\!\!\!\!\!\!\\
 & \multicolumn{3}{c}{LRT} & \multicolumn{2}{c}{Fuji} & \multicolumn{2}{c}{Schott} & \multicolumn{2}{c}{ChenQin} & \multicolumn{2}{c}{Zhang} & \!\!\!\!\!\!\!\\[-11pt]
& \multicolumn{3}{c}{\rule{105pt}{.5pt}} & \multicolumn{2}{c}{\rule{69pt}{.5pt}} & \multicolumn{2}{c}{\rule{69pt}{.5pt}} & \multicolumn{2}{c}{\rule{69pt}{.5pt}} & \multicolumn{2}{c}{\rule{69pt}{.5pt}} & \!\!\!\!\!\!\!\\[-4pt]
$\alpha$ & exact & asymp & simul & asymp & simul & asymp & simul & asymp & simul & asymp & simul & \!\!\!\!\!\!\!\\
\hline\\[-8pt]
0.01 & 0.137 & 0.205 & 0.140 & 0.271 & 0.147 & 0.267 & 0.147 & 0.055 & 0.127 & 0.232 & 0.171 \\
0.05 & 0.344 & 0.383 & 0.346 & 0.438 & 0.363 & 0.433 & 0.363 & 0.141 & 0.300 & 0.432 & 0.396
\end{tabular}
\end{table}

\vspace{-.5cm}

\begin{table}[ht!]
\centering
\caption{\label{table34n}Skew Normal distribution with circular covariance matrix with correlations 0.108 to 0.216 and $n_k=4,8$\\[-9pt]}
\begin{tabular}{ccccccc}
\multicolumn{7}{c}{$\boldsymbol H_0$}\\
 & \multicolumn{2}{c}{LRT} & & & \\[-1pt]
\cline{2-3}\\[-19pt]
$\alpha$ & exact & asymp & \raisebox{5pt}{Fuji} & \raisebox{5pt}{Schott} & \raisebox{5pt}{ChenQin} & \raisebox{5pt}{Zhang}\\
\hline
0.01 & 0.010 & 0.020 & 0.029 & 0.028 & 0.002 & 0.023 \\
0.05 & 0.050 & 0.062 & 0.073 & 0.072 & 0.012 & 0.073 \\[7pt]
\end{tabular}
\begin{tabular}{@{\extracolsep{-2pt}}ccccccccccccc}
\multicolumn{12}{c}{~~~~~~~~~~~~~~~~$\boldsymbol H_1$ (shift=+0.75,--0.75)} & \!\!\!\!\!\!\!\\
 & \multicolumn{3}{c}{LRT} & \multicolumn{2}{c}{Fuji} & \multicolumn{2}{c}{Schott} & \multicolumn{2}{c}{ChenQin} & \multicolumn{2}{c}{Zhang} & \!\!\!\!\!\!\!\\[-11pt]
& \multicolumn{3}{c}{\rule{105pt}{.5pt}} & \multicolumn{2}{c}{\rule{69pt}{.5pt}} & \multicolumn{2}{c}{\rule{69pt}{.5pt}} & \multicolumn{2}{c}{\rule{69pt}{.5pt}} & \multicolumn{2}{c}{\rule{69pt}{.5pt}} & \!\!\!\!\!\!\!\\[-4pt]
$\alpha$ & exact & asymp & simul & asymp & simul & asymp & simul & asymp & simul & asymp & simul & \!\!\!\!\!\!\!\\
\hline\\[-8pt]
0.01 & 0.192 & 0.272 & 0.190 & 0.264 & 0.141 & 0.260 & 0.141 & 0.049 & 0.125 & 0.223 & 0.162 \\
0.05 & 0.429 & 0.469 & 0.428 & 0.442 & 0.361 & 0.438 & 0.361 & 0.142 & 0.335 & 0.435 & 0.397
\end{tabular}
\end{table}

\vspace{-.5cm}

\begin{table}[ht!]
\centering
\caption{\label{table35n}Skew Normal distribution with circular covariance matrix with correlations 0.108 to 0.216 and $n_k=4,8,12$\\[-9pt]}
\begin{tabular}{ccccccc}
\multicolumn{7}{c}{$\boldsymbol H_0$}\\
 & \multicolumn{2}{c}{LRT} & & & \\[-1pt]
\cline{2-3}\\[-19pt]
$\alpha$ & exact & asymp & \raisebox{5pt}{Fuji} & \raisebox{5pt}{Schott} & \raisebox{5pt}{ChenQin} & \raisebox{5pt}{Zhang}\\
\hline
0.01 & 0.011 & 0.018 & 0.025 & 0.023 & --- & --- \\
0.05 & 0.051 & 0.060 & 0.070 & 0.066 & --- & --- \\[7pt]
\end{tabular}
\begin{tabular}{@{\extracolsep{-2pt}}ccccccccccccc}
\multicolumn{12}{c}{~~~~~~~~~~~~~~~~$\boldsymbol H_1$ (shift=+0.75,+0.75)} & \!\!\!\!\!\!\!\\
 & \multicolumn{3}{c}{LRT} & \multicolumn{2}{c}{Fuji} & \multicolumn{2}{c}{Schott} & \multicolumn{2}{c}{ChenQin} & \multicolumn{2}{c}{Zhang} & \!\!\!\!\!\!\!\\[-11pt]
& \multicolumn{3}{c}{\rule{105pt}{.5pt}} & \multicolumn{2}{c}{\rule{69pt}{.5pt}} & \multicolumn{2}{c}{\rule{69pt}{.5pt}} & \multicolumn{2}{c}{\rule{69pt}{.5pt}} & \multicolumn{2}{c}{\rule{69pt}{.5pt}} & \!\!\!\!\!\!\!\\[-4pt]
$\alpha$ & exact & asymp & simul & asymp & simul & asymp & simul & asymp & simul & asymp & simul & \!\!\!\!\!\!\!\\
\hline\\[-8pt]
0.01 & 0.361 & 0.438 & 0.352 & 0.514 & 0.370 & 0.498 & 0.370 & --- & --- & --- & --- \\
0.05 & 0.627 & 0.658 & 0.623 & 0.702 & 0.639 & 0.692 & 0.639 & --- & --- & --- & ---
\end{tabular}
\end{table}

\vspace{-.5cm}

\begin{table}[ht!]
\centering
\caption{\label{table36n}Skew Normal distribution with circular covariance matrix with correlations 0.108 to 0.216 and $n_k=4,8,12$\\[-9pt]}
\begin{tabular}{ccccccc}
\multicolumn{7}{c}{$\boldsymbol H_0$}\\
 & \multicolumn{2}{c}{LRT} & & & \\[-1pt]
\cline{2-3}\\[-19pt]
$\alpha$ & exact & asymp & \raisebox{5pt}{Fuji} & \raisebox{5pt}{Schott} & \raisebox{5pt}{ChenQin} & \raisebox{5pt}{Zhang}\\
\hline
0.01 & 0.010 & 0.017 & 0.024 & 0.022 & --- & --- \\
0.05 & 0.051 & 0.059 & 0.069 & 0.065 & --- & --- \\[7pt]
\end{tabular}
\begin{tabular}{@{\extracolsep{-2pt}}ccccccccccccc}
\multicolumn{12}{c}{~~~~~~~~~~~~~~~~$\boldsymbol H_1$ (shift=+0.75,--0.75)} & \!\!\!\!\!\!\!\\
 & \multicolumn{3}{c}{LRT} & \multicolumn{2}{c}{Fuji} & \multicolumn{2}{c}{Schott} & \multicolumn{2}{c}{ChenQin} & \multicolumn{2}{c}{Zhang} & \!\!\!\!\!\!\!\\[-11pt]
& \multicolumn{3}{c}{\rule{105pt}{.5pt}} & \multicolumn{2}{c}{\rule{69pt}{.5pt}} & \multicolumn{2}{c}{\rule{69pt}{.5pt}} & \multicolumn{2}{c}{\rule{69pt}{.5pt}} & \multicolumn{2}{c}{\rule{69pt}{.5pt}} & \!\!\!\!\!\!\!\\[-4pt]
$\alpha$ & exact & asymp & simul & asymp & simul & asymp & simul & asymp & simul & asymp & simul & \!\!\!\!\!\!\!\\
\hline\\[-8pt]
0.01 & 0.494 & 0.573 & 0.496 & 0.517 & 0.371 & 0.501 & 0.371 & --- & --- & --- & --- \\
0.05 & 0.748 & 0.772 & 0.746 & 0.717 & 0.654 & 0.706 & 0.654 & --- & --- & --- & ---
\end{tabular}
\end{table}

\vspace{-.5cm}

\begin{table}[ht!]
\centering
\caption{\label{table37n}Skew T distribution with 5 d.f.\ and with circular covariance matrix with correlations 0.108 to 0.216 and $n_k=4,4$\\[-9pt]}
%
\end{table}

\vspace{-.5cm}

\begin{table}[ht!]
\centering
\caption{\label{table38n}Skew T distribution with 5 d.f.\ and with circular covariance matrix with correlations 0.108 to 0.216 and $n_k=4,4$\\[-9pt]}
%
\end{table}

\vspace{-.5cm}

\begin{table}[ht!]
\centering
\caption{\label{table39n}Skew T distribution with 5 d.f.\ and with circular covariance matrix with correlations 0.108 to 0.216 and $n_k=4,8$\\[-9pt]}
%
\end{table}

\vspace{-.5cm}

\begin{table}[ht!]
\centering
\caption{\label{table41n}Skew T distribution with 5 d.f.\ and with circular covariance matrix with correlations 0.108 to 0.216 and $n_k=4,8,12$\\[-9pt]}
\begin{tabular}{ccccccc}
\multicolumn{7}{c}{$\boldsymbol H_0$}\\
 & \multicolumn{2}{c}{LRT} & & & \\[-1pt]
\cline{2-3}\\[-19pt]
$\alpha$ & exact & asymp & \raisebox{5pt}{Fuji} & \raisebox{5pt}{Schott} & \raisebox{5pt}{ChenQin} & \raisebox{5pt}{Zhang}\\
\hline
0.01 & 0.033 & 0.044 & 0.024 & 0.022 & --- & --- \\
0.05 & 0.088 & 0.097 & 0.064 & 0.060 & --- & --- \\[7pt]
\end{tabular}%
\end{table}

\vspace{-.5cm}

\begin{table}[ht!]
\centering
\caption{\label{table42n}Skew T distribution with 5 d.f.\ and with circular covariance matrix with correlations 0.108 to 0.216 and $n_k=4,8,12$\\[-9pt]}
%
\end{table}

\vspace{-.5cm}

\begin{table}[ht!]
\centering
\caption{\label{table47n}Skew T distribution with 3 d.f.\ and with circular covariance matrix with correlations 0.108 to 0.216 and $n_k=4,8,12$\\[-9pt]}
%
\end{table}

\clearpage

\begin{table}[ht!]
\centering
\caption{\label{table49n}Skew Cauchy distribution with circular covariance matrix with correlations 0.108 to 0.216 and $n_k=4,4$\\[-9pt]}
%
\end{table}

\vspace{-.5cm}

\begin{table}[ht!]
\centering
\caption{\label{table50n}Skew Cauchy distribution with circular covariance matrix with correlations 0.108 to 0.216 and $n_k=4,4$\\[-9pt]}
%
\end{table}

\vspace{-.5cm}

\begin{table}[ht!]
\centering
\caption{\label{table51n}Skew Cauchy distribution with circular covariance matrix with correlations 0.108 to 0.216 and $n_k=4,8$\\[-9pt]}
%
\end{table}

\vspace{-.5cm}

\begin{table}[ht!]
\centering
\caption{\label{table52n}Skew Cauchy distribution with circular covariance matrix with correlations 0.108 to 0.216 and $n_k=4,8$\\[-9pt]}
%
\end{table}

\vspace{-.5cm}

\begin{table}[ht!]
\centering
\caption{\label{table53n}Skew Cauchy distribution with circular covariance matrix with correlations 0.108 to 0.216 and $n_k=4,8,12$\\[-9pt]}
%
\end{table}

\vspace{-.5cm}

\begin{table}[ht!]
\centering
\caption{\label{table54n}Skew Cauchy distribution with circular covariance matrix with correlations 0.108 to 0.216 and $n_k=4,8,12$\\[-9pt]}
%
\end{table}

\pagebreak






\centerline{*************************************************}

\begin{table}[t]
\centering
\caption{\label{table55n}Normal distribution with circular covariance matrix with correlations 0.250 to 0.350 and $n_k=4,4$\\[-9pt]}
%
\end{table}

\vspace{-.5cm}

\begin{table}[ht!]
\centering
\caption{\label{table67n}T distribution with 5 d.f.\ with circular covariance matrix with correlations 0.250 to 0.350 and $n_k=4,4$\\[-9pt]}
%
\end{table}

\vspace{-.5cm}

\begin{table}[ht!]
\centering
\caption{\label{table68n}T distribution with 5 d.f.\ with circular covariance matrix with correlations 0.250 to 0.350 and $n_k=4,4$\\[-9pt]}
%
\end{table}

\vspace{-.5cm}

\begin{table}[ht!]
\centering
\caption{\label{table69n}T distribution with 5 d.f.\ with circular covariance matrix with correlations 0.250 to 0.350 and $n_k=4,8$\\[-9pt]}
%
\end{table}

\vspace{-.5cm}

\begin{table}[ht!]
\centering
\caption{\label{table71n}T distribution with 5 d.f.\ with circular covariance matrix with correlations 0.250 to 0.350 and $n_k=4,8,12$\\[-9pt]}
%
\end{table}

\vspace{-.5cm}

\begin{table}[ht!]
\centering
\caption{\label{table72n}T distribution with 5 d.f.\ with circular covariance matrix with correlations 0.250 to 0.350 and $n_k=4,8,12$\\[-9pt]}
%
\end{table}

\vspace{-.5cm}

\begin{table}[ht!]
\centering
\caption{\label{table77n}T distribution with 3 d.f.\ with circular covariance matrix with correlations 0.250 to 0.350 and $n_k=4,8,12$\\[-9pt]}
%
\end{table}

\vspace{-.5cm}

\begin{table}[ht!]
\centering
\caption{\label{table83n}Cauchy distribution with circular covariance matrix with correlations 0.250 to 0.350 and $n_k=4,8,12$\\[-9pt]}
%
\end{table}

\vspace{-.5cm}

\begin{table}[ht!]
\centering
\caption{\label{table89n}Skew Normal distribution with circular covariance matrix with correlations 0.250 to 0.350 and $n_k=4,8,12$\\[-9pt]}
%
\end{table}

\clearpage

\vspace{-.5cm}

\begin{table}[ht!]
\centering
\caption{\label{table103n}Skew Cauchy distribution with circular covariance matrix with correlations 0.250 to 0.350 and $n_k=4,4$\\[-9pt]}
%
\end{table}

\vspace{-.5cm}

\begin{table}[ht!]
\centering
\caption{\label{table107n}Skew Cauchy distribution with circular covariance matrix with correlations 0.250 to 0.350 and $n_k=4,8,12$\\[-9pt]}
%
\end{table}

\vspace{-.5cm}

\clearpage

\centerline{*************************************************}

\begin{table}[ht!]
\centering
\caption{\label{table109n}Normal distribution with circular covariance matrix with correlations 0.350 to 0.490 and $n_k=4,4$\\[-9pt]}
%
\end{table}

\vspace{-.5cm}

\begin{table}[ht!]
\centering
\caption{\label{table125n}T distribution with 5 d.f.\ and with circular covariance matrix with correlations 0.350 to 0.490 and $n_k=4,8$\\[-9pt]}
\end{table}


\clearpage


\begin{table}[ht!]
\centering
\caption{\label{table144n}Skew Normal distribution with circular covariance matrix with correlations 0.350 to 0.490 and $n_k=4,4$\\[-9pt]}
%
\end{table} 

\vspace{-.5cm}

\clearpage

\centerline{************************************************}


\begin{table}[ht!]
\centering
\caption{\label{table178n}Normal distribution with circular covariance matrix with correlations 0.358 to 0.501 and $n_k=4,4$\\[-9pt]}
%
\end{table} 

\clearpage
\vspace{-.5cm}

\begin{table}[ht!]
\centering
\caption{\label{table221n}Skew T distribution with 5 d.f.\ and with circular covariance matrix with correlations 0.358 to 0.501 and $n_k=4,8,12$\\[-9pt]}
%
\end{table} 

\clearpage

\vspace{-.5cm}

\centerline{************************************}

\begin{table}[ht!]
\centering
\caption{\label{table238n}Normal distribution with circular covariance matrix with correlations 0.500 to 0.600 and $n_k=4,4$\\[-9pt]}
%
\end{table} 

\clearpage

\vspace{-.5cm}

\begin{table}[ht!]
\centering
\caption{\label{table286n}Skew T distribution with 5 d.f.\ and with circular covariance matrix with correlations 0.500 to 0.600 and $n_k=4,4$\\[-9pt]}
%
\end{table}

\clearpage

\centerline{**************************************}

\begin{table}[ht!]
\centering
\caption{\label{table307n}Normal distribution with circular covariance matrix with correlations 0.606 to 0.697 and $n_k=4,4$\\[-9pt]}
%
\end{table}

\clearpage
\vspace{-.5cm}

\begin{table}[ht!]
\centering
\caption{\label{table355n}Skew T distribution with 5 d.f.\ and with circular covariance matrix with correlations 0.606 to 0.697 and $n_k=4,8$\\[-9pt]}
%
\end{table}

\clearpage

\vspace{-.5cm}

\centerline{*******************************}

\begin{table}[ht!]
\centering
\caption{\label{table378n}Normal distribution with positive-definite covariance matrix with correlations 0.123 to 0.329 and $n_k=4,4$\\[-9pt]}
%
\end{table}

\vspace{-.5cm}

\begin{table}[ht!]
\centering
\caption{\label{table379n}Normal distribution with positive-definite covariance matrix with correlations 0.123 to 0.329 and $n_k=4,4$\\[-9pt]}
%
\end{table}

\vspace{-.5cm}

\begin{table}[ht!]
\centering
\caption{\label{table380n}Normal distribution with positive-definite covariance matrix with correlations 0.123 to 0.329 and $n_k=4,4$\\[-9pt]}
%
\end{table}

\vspace{-.5cm}

\begin{table}[ht!]
\centering
\caption{\label{table381n}Normal distribution with positive-definite covariance matrix with correlations 0.123 to 0.329 and $n_k=4,4$\\[-9pt]}
%
\end{table}

\clearpage
\vspace{-.5cm}

\begin{table}[ht!]
\centering
\caption{\label{table384n}Normal distribution with positive-definite covariance matrix with correlations 0.123 to 0.329 and $n_k=4,8$\\[-9pt]}
%
\end{table}

\vspace{-.5cm}

\begin{table}[ht!]
\centering
\caption{\label{table386n}Normal distribution with positive-definite covariance matrix with correlations 0.123 to 0.329 and $n_k=4,8,12$\\[-9pt]}
%
\end{table}

\vspace{-.5cm}

\begin{table}[ht!]
\centering
\caption{\label{table387n}Normal distribution with positive-definite covariance matrix with correlations 0.123 to 0.329 and $n_k=4,8,12$\\[-9pt]}
%
\end{table}

\vspace{-.5cm}

\begin{table}[ht!]
\centering
\caption{\label{table388n}Normal distribution with positive-definite covariance matrix with correlations 0.123 to 0.329 and $n_k=4,8,12$\\[-9pt]}
%
\end{table}

\vspace{-.5cm}

\begin{table}[ht!]
\centering
\caption{\label{table390n}T distribution with 5 d.f.\ and with positive-definite covariance matrix with correlations 0.123 to 0.329 and $n_k=4,4$\\[-9pt]}
%
\end{table}

\vspace{-.5cm}

\begin{table}[ht!]
\centering
\caption{\label{table391n}T distribution with 5 d.f.\ and with positive-definite covariance matrix with correlations 0.123 to 0.329 and $n_k=4,4$\\[-9pt]}
%
\end{table}

\vspace{-.5cm}

\begin{table}[ht!]
\centering
\caption{\label{table392n}T distribution with 5 d.f.\ and with positive-definite covariance matrix with correlations 0.123 to 0.329 and $n_k=4,8$\\[-9pt]}
%
\end{table}

\vspace{-.5cm}

\begin{table}[ht!]
\centering
\caption{\label{table394n}T distribution with 5 d.f.\ and with positive-definite covariance matrix with correlations 0.123 to 0.329 and $n_k=4,8,12$\\[-9pt]}
%
\end{table}

\vspace{-.5cm}

\begin{table}[ht!]
\centering
\caption{\label{table395n}T distribution with 5 d.f.\ and with positive-definite covariance matrix with correlations 0.123 to 0.329 and $n_k=4,8,12$\\[-9pt]}
%
\end{table}

\vspace{-.5cm}

\begin{table}[ht!]
\centering
\caption{\label{table400n}T distribution with 3 d.f.\ and with positive-definite covariance matrix with correlations 0.123 to 0.329 and $n_k=4,8,12$\\[-9pt]}
%
\end{table}

\vspace{-.5cm}

\clearpage
\pagebreak

\bibliographystyle{Chicago}
\bibliography{bibliography}

\end{document}